\documentclass[twocolumn]{aastex631}

\shorttitle{Revisiting the TTV Planets}
\shortauthors{Lammers \& Winn}

\graphicspath{{./}{}}
\usepackage{amsmath}
\usepackage{comment}

\begin{document}

\title{TTV-Not-So-Fast: Uniqueness and Degeneracy in Perturbing Planet Parameters}

\author[0000-0001-9985-0643]{Caleb Lammers}
\affiliation{Department of Astrophysical Sciences, Princeton University, 4 Ivy Lane, Princeton, NJ 08544, USA}

\author[0000-0002-4265-047X]{Joshua N.\ Winn}
\affiliation{Department of Astrophysical Sciences, Princeton University, 4 Ivy Lane, Princeton, NJ 08544, USA}

\begin{abstract}

Nontransiting planets can reveal themselves through transit timing variations (TTVs), but inferring the properties of the perturbing planet is a highly degenerate inverse problem. We present a systematic reassessment of all $12$ published cases in which a nontransiting planet was claimed to have been uniquely characterized using TTVs. Two systems (KOI-142 and Kepler-419) stand out clearly with compelling evidence for unique solutions. Two other systems (KOI-872 and KOI-884) exhibit complex degeneracies, but the data are just precise enough to single out a best solution. Six systems (Kepler-82, Kepler-411, Kepler-725, KOI-134, Kepler-138, and TOI-4562) admit multiple viable solutions involving very different perturbing planets. In the remaining two systems (WASP-18 and WASP-126), the evidence for any perturbing planet is weak. We find that a necessary (but not sufficient) condition for a unique solution is the detection of short-timescale TTV structure associated with conjunctions, either in the near-resonant ``chopping'' regime or in eccentric systems with phase-dependent close approaches. In some systems, aliasing of the synodic period leads to ambiguities in associating observed TTV timescales with physical timescales, threatening uniqueness. Our results highlight the difficulty of achieving unique solutions in TTV inversions and underscore the need for long time baselines, accurate timing uncertainties, and complementary constraints from radial velocities or other observations when characterizing nontransiting planets.

\end{abstract}

\section{Introduction}
\label{sec:intro}

In an idealized two-body star-planet system, the time interval between successive planetary transits would be exactly constant. Gravitational perturbations from additional, unseen planets cause departures from periodicity known as transit timing variations (TTVs). More than two decades ago, \citet{Holman&Murray2005} and \citet{Agol2005} recognized that transit timing provides a pathway to the discovery of nontransiting planets, in some cases with sensitivity to lower-mass planets than other detection techniques. Observational searches for TTVs were undertaken shortly thereafter, mainly targeting hot Jupiters and yielding null results \citep[e.g.,][]{Steffen&Agol2005, MillerRicci2008, Agol2009}. This is now interpreted as evidence that hot Jupiters are rarely accompanied by nearby planets \citep{Steffen2012b, Hord2021, Sha+2026}.

The transit-timing method reached maturity with NASA's {\it Kepler} mission \citep{Borucki2010}, whose continuous high-precision photometry provided precise transit times for many systems of small planets. In some cases, the mere existence of TTVs was used to confirm {\it Kepler} planet candidates \citep[e.g.,][]{Holman2010, Fabrycky2012, Ford2012b, Steffen2013}. In systems where the perturbing planets are also transiting, TTV analyses constrained the planets' masses and eccentricities \citep[e.g.,][]{Wu&Lithwick2013, Hadden&Lithwick2014, Hadden&Lithwick2017}. More challenging to interpret were systems in which the observed TTVs of one planet are caused by an otherwise undetected planet. For example, \citet{Ballard2011} measured the sinusoidally varying TTVs of Kepler-19\,b with high statistical significance, but found that this trend could be produced by planets with masses, periods, and eccentricities drawn from different families of solutions. This was the earliest demonstration on real data of the difficulty of finding a unique solution to the inverse TTV problem, by which we mean breaking the discrete degeneracies and narrowing down the perturber's properties to a singly connected region of parameter space.

A sinusoidal TTV pattern is insufficient for a unique solution because a perturber near a $j$:$j\,{-}\,k$ mean-motion resonance (MMR) with the transiting planet will induce nearly sinusoidal TTVs with a characteristic ``super-period''
\begin{equation}
\label{eqn:TTV_super_period}
P_\mathrm{TTV} = \frac{1}{j}\frac{P_2}{|\Delta|}~,~\mathrm{where}~ \Delta = \frac{j - k}{j} \frac{P_2}{P_1} - 1
\end{equation}
measures the fractional departure from exact commensurability. Here, $P_2$ is the orbital period of the exterior perturber and $P_1$ is that of the transiting planet. The super-period is the timescale over which the longitude of conjunction advances by $2\pi/k$ radians relative to exact resonance, at which point the conjunction geometry repeats. For first-order MMRs ($k\,{=}\,1$) and nearly circular orbits, the associated TTV amplitude is proportional to $(P_1/\Delta)(m_2/M_\star)$, where $m_2$ and $M_\star$ are the masses of the perturber and the host star \citep{Agol2005, Lithwick2012}. In general, the amplitude also depends sensitively on the planets' eccentricities and mutual inclination \citep{Nesvorny&Morbidelli2008, Nesvorny2009, Lithwick2012}. Perturbers near different MMRs, with different masses and orbital elements, can therefore induce sinusoidal TTVs with indistinguishable periods and amplitudes.

In the case of Kepler-19\,b, the degeneracies were eventually broken using radial velocity (RV) data \citep{Malavolta2017}. The first unique solution of the TTV inverse problem without any auxiliary data was achieved by \citet{Nesvorny2012} for KOI-872. Based on TTVs and TDVs (transit duration variations), they showed that the data could only be explained by a perturbing planet of mass ${\approx}\,0.4~M_\mathrm{J}$ and period ${\approx}\,57$~days, near the $5$:$3$ MMR with the transiting planet. As more systems were investigated, a practical rule of thumb emerged: unique solutions generally require the detection of variations on timescales shorter than the super-period $P_\mathrm{TTV}$, typically associated with the synodic (conjunction) period \citep{Nesvorny&Vokrouhlicky2014, Deck&Agol2015}. These faster ``chopping'' variations tend to have lower amplitudes than the long-period $P_\mathrm{TTV}$ signal, and are therefore more difficult to detect. Consequently, only datasets with high timing precision and long temporal baselines are likely to resolve both the super-period and the conjunction-scale structure needed for a unique solution.

According to the NASA Exoplanet Archive \citep{Christiansen2025},\footnote{\url{https://exoplanetarchive.ipac.caltech.edu}} there are currently $12$ planets that were detected, and for which a unique solution was achieved, using only transit-timing data. Table~\ref{table:names} provides the planets' names and citations. We were motivated to revisit these ``TTV planets'' because several recent claims of uniqueness were based on datasets that  show no clear evidence of conjunction-scale signals, seemingly at odds with the rule of thumb described above. After our preliminary reanalysis of a few of these systems revealed that the published solutions were not unique, we decided to conduct a systematic audit of all $12$ cases to clarify the conditions that allow TTV solutions to be considered unique.

Section~\ref{sec:TTV_modeling} describes our dynamical modeling and strategy for searching for possible TTV solutions. Section~\ref{sec:indiv_TTV_results} presents the results for each system. Section~\ref{sec:discussion} synthesizes these findings and draws some general lessons about degeneracies and uniqueness in TTV inversions.

\begin{table}
    \centering
    \caption{TTV Planets with Reported Unique Solutions}
    \begin{tabular}{c|l}
     \hline
    Name & Reference \\
     \hline
     KOI-872\,c & \citet{Nesvorny2012}\\
     KOI-142\,c & \citet{Nesvorny2013}\\
     KOI-884\,e & \citet{Nesvorny2014}\\
     Kepler-419\,c & \citet{Dawson2014}\\
     Kepler-82\,f & \citet{Freudenthal2019}\\
     WASP-18\,c & \citet{Pearson2019}\\
     WASP-126\,c & \citet{Pearson2019}\\
     Kepler-411\,e & \citet{Sun2019}\\
     Kepler-138\,e & \citet{Piaulet2023}\\
     TOI-4562\,c & \citet{Fermiano2024}\\
     KOI-134\,c & \citet{Nabbie2025}\\
     Kepler-725\,c & \citet{Sun2025}\\
     \hline
    \end{tabular}
    \label{table:names}
\end{table}

\section{TTV modeling}
\label{sec:TTV_modeling}

\subsection{Calculating TTVs and TDVs}
\label{sec:TTVfast}

We modeled TTVs and TDVs using $N$-body simulations performed with the \texttt{TTVFast} code \citep{Deck2014}.\footnote{We used the Python wrapper of \texttt{TTVFast} available at \url{https://github.com/simonrw/ttvfast-python}.} \texttt{TTVFast} takes advantage of an efficient \citet{Wisdom&Holman1991} integrator to advance the positions of the planets and computes transit times by assuming Keplerian motion between the two time steps bracketing each transit. To ensure high accuracy, we set the Wisdom-Holman time step to be $5$\% of the minimum periastron passage timescale across all planets, calculated as \citep{Wisdom2015}
\begin{equation}
    T_\mathrm{peri} = \frac{2\pi}{\dot{f}_\mathrm{peri}} = P\,\frac{(1 - e)^{3/2}}{(1 + e)^{1/2}}~.
\end{equation}

The model included seven free parameters per planet: mass $m$, period $P$, mean anomaly $M$, inclination $i$, longitude of ascending node $\Omega$, and eccentricity vector components $e \sin \omega$ and $e \cos \omega$. Orbital elements are the osculating elements at the reference epoch $2{,}454{,}900$\,BJD.\footnote{The reference epoch is March 3, 2009, 12:00~UT, a few days before the launch of the {\it Kepler} telescope.}
For transiting planets, instead of the inclination, we used the impact parameter
\begin{equation}
\label{eqn:impact_param}
b = \frac{a \cos i}{R_\star} \left(\frac{1 - e^2}{1 + e \sin \omega}\right)~.
\end{equation}
We set $\Omega\,{=}\,0^\circ$ for the innermost transiting planet, without loss of generality. Transit durations were estimated as
\begin{equation}
\label{eqn:transit_durations}
T_\mathrm{dur} = \frac{2 R_\star}{v_\mathrm{sky}} \sqrt{1 - b^2 }~,
\end{equation}
where $R_\star$ is the star's radius, $b$ is the impact parameter, and $v_\mathrm{sky}$ is the sky-projected velocity of the planet, relative to the star, evaluated at the transit midpoint. For reproducibility, the text reports the values we adopted for each system's stellar mass and radius.

\subsection{TTV search strategy}
\label{sec:TTV_search}

Characterizing the posterior distribution of TTV models is challenging because the parameter space is high-dimensional and the likelihood surface is typically multi-modal. Viable solutions often occupy narrow, disjoint regions of parameter space. A variety of approaches have been used to tackle this problem, including nested sampling \citep[e.g.,][]{Nesvorny2012, Nesvorny2013, Nesvorny2014}, Markov Chain Monte Carlo (MCMC) methods \citep{Dawson2014}, genetic algorithms and differential evolution codes \citep{Sun2019, Sun2025}, and grid searches supplemented with MCMC \citep{Nabbie2025}.

Rather than attempting to fully characterize the posterior distribution, we adopted a targeted search strategy designed to identify plausible families of solutions. Our approach emphasizes dense grids of initial conditions close to MMRs, combined with large-scale local optimization. For each system, we devoted at least $10^4$ CPU hours to the search process, typically entailing ${\sim}\,10^6$ independent optimizations. Below, we describe the generation of initial conditions and the optimization process.

For transiting planets, initial guesses for the masses were drawn from the mass-radius relation of \citet{Chen2017} based on the measured radii. Masses exceeding $10~M_\mathrm{J}$ were rejected and redrawn. To focus the search on low-eccentricity configurations, which are known to be common for {\it Kepler} systems \citep[e.g.,][]{VanEylen&Albrecht2015, Hadden&Lithwick2017}, we sampled $e \cos \omega$ and $e \sin \omega$ independently from $\mathcal{U}(-0.1,~0.1)$. During optimization, however, masses and eccentricities were allowed to vary freely beyond these initial ranges. The impact parameter of the innermost transiting planet was sampled from $\mathcal{U}(0,\,1)$, and additional transiting planets were assigned impact parameters from $\mathcal{U}(-1,\,1)$. Orbital periods were drawn from Gaussian distributions centered on values obtained from constant-period fits, with a standard deviation of $0.1$~day (intentionally broader than the formal uncertainties, to allow exploration). The initial mean anomalies were chosen such that, in the absence of planet-planet interactions, the first transit midpoint would correspond exactly to inferior conjunction.

For the perturbing planet, initial masses were drawn from a log-uniform distribution spanning from $10~M_\oplus$ to $10~M_\mathrm{J}$. The eccentricity vector components were sampled from $\mathcal{U}(-0.1,~0.1)$, and the mean anomaly at the reference epoch was drawn from $\mathcal{U}(0^\circ,~360^\circ)$. To concentrate on approximately coplanar configurations, we sampled a mutual inclination $i_\mathrm{mut}$ from $\mathcal{U}(0^\circ,~30^\circ)$ and an azimuthal angle around the transiting planet's orbit normal from $\mathcal{U}(0^\circ,~360^\circ)$. The orbital inclination and longitude of the ascending node
were computed from these angles.

The initial orbital period of the perturbing planet was restricted to values near MMRs with the transiting planet. We focused on the twenty $j$:$j\,{-}\,k$ MMRs with period ratios between $1.15$ and $10$ satisfying $k\,{=}\,1$ (first order), $k\,{=}\,2$ (second order), or $j\,{-}\,k\,{=}\,1$. This set of MMRs was chosen based on the published TTV planets and the requirement that MMRs be sufficiently strong to generate large TTV amplitudes. To assign the perturber's orbital period, a resonance was first selected at random from the set of twenty possibilities. Then, we used the observed TTV timescale ($P_\mathrm{obs}$), determined via a sinusoidal fit, to sample a resonance offset parameter $\Delta$ from $\mathcal{U}(\Delta_\mathrm{min},\,\Delta_\mathrm{max})$, where the bounds ensure that the TTV super-period (Equation~\ref{eqn:TTV_super_period}) is between $0.5 P_\mathrm{obs}$ and $2 P_\mathrm{obs}$. The resulting period ratio is
\begin{equation} 
\label{eqn:P2}
\mathcal{P} = (1 \pm \Delta)\left(\frac{j}{j - k}\right)~.
\end{equation}
Planets were assigned with equal probability to be wide ($\Delta\,{>}\,0$) or narrow ($\Delta\,{<}\,0$) of resonance, and to be interior or exterior to the transiting planet. Although initial periods were clustered near resonances, they were allowed to vary freely during optimization and explore solutions near higher-order resonances or far from any meaningful resonances. By limiting the initial guesses to a plausible set of resonances and $\Delta$ values, computational effort could be devoted to the most promising regions of parameter space.

Each initial condition was passed to a local optimizer tasked with minimizing the sum of squared residuals ($\chi^2$) of the transit time and duration measurements. We tested several algorithms implemented in \texttt{scipy} \citep{Virtanen2020}, including Nelder-Mead, L-BFGS-B, Powell, COBYLA, Dog Leg, Levenberg-Marquardt, and the Trust Region Framework (TRF). We obtained the most reliable convergence with TRF, as implemented in the \texttt{optimize.least\_squares} class. TRF is a Gauss-Newton method that, at each iteration, minimizes a quadratic model of the residuals within an adaptively adjusted ``trust region,'' while enforcing parameter bounds via reflective steps. The adaptive control of the trust-region radius appears particularly well suited to TTV fitting, where solutions occupy small and irregular regions of parameter space. For each of the ${\sim}\,10^6$ initial conditions, we performed three TRF optimization passes. We validated many of the results presented below by repeating selected searches with the Nelder-Mead algorithm implemented in \texttt{optimize.minimize}, obtaining consistent solutions.

Some previous TTV analyses have discarded solutions that fit the data but are dynamically unstable over long timescales (see, e.g., \citealt{Sun2019}, \citealt{Sun2025}, and \citealt{Nabbie2025}). We did not adopt this approach because stable and unstable solutions are often tightly interwoven in parameter space. Small changes to orbital periods and eccentricities could convert an unstable configuration into a stable one without significantly degrading the quality of the fit. For resonant systems, several authors have shown that initial conditions drawn directly from observational posteriors can be overwhelmingly unstable, and that additional model ingredients (such as convergent migration) are needed to locate the small regions of stable solutions \citep[see, e.g.,][]{Tamayo2017, Lammers&Winn2024, Lu2025}. Although we did not discard solutions based on stability during the search, period ratios where solutions are likely to be unstable are highlighted in red in the figures below. Stability boundaries were estimated using the \citet{Gladman1993} criterion, assuming circular orbits, massless companions, and transiting-planet masses from the \citet{Chen2017} mass-radius relation. The solutions highlighted in the text and tables seem plausibly long-term stable.

\begin{figure*}
\centering
\includegraphics[width=0.95\textwidth]{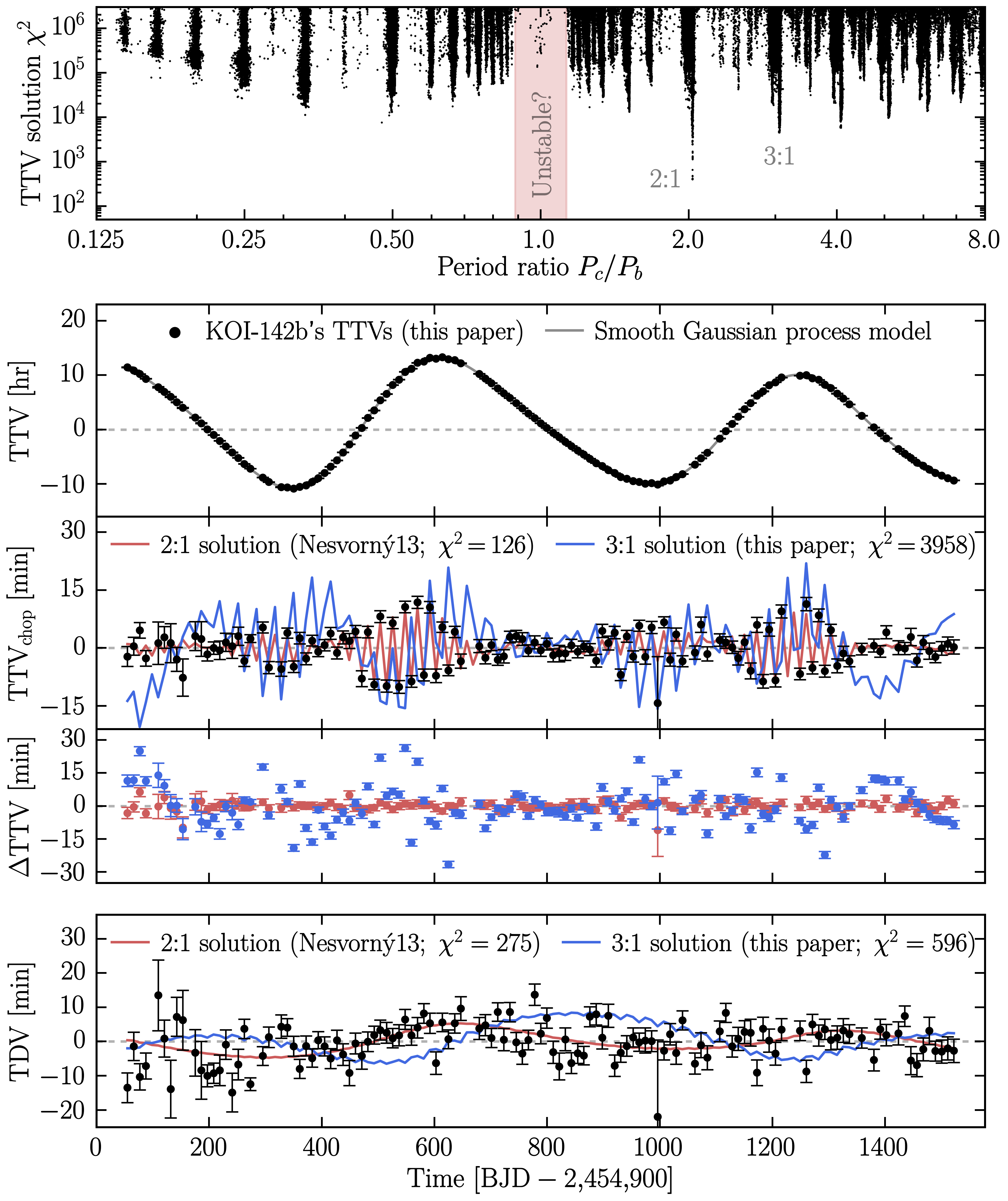}
\caption{{\bf KOI-142}. {\it Top panel:} $\chi^2$ versus period ratio ($N_\mathrm{dof}\,{=}\,231$). The minimum near the $2$:$1$ MMR is substantially deeper than any competing solution. {\it Second panel:} Measured TTVs from this work (Section~\ref{sec:KOI142}). The gray curve shows a Gaussian process (GP) regression used solely as a smooth interpolant to separate long- and short-timescale variations. {\it Third panel:} High-pass-filtered TTVs obtained by subtracting the GP. The red curve shows the best variant of the $2$:$1$ solution \citep{Nesvorny2013}; the blue curve shows the $3$:$1$ solution, which is second-best. The short-timescale ``chopping'' signal strongly favors the $2$:$1$ model. {\it Fourth panel:} Residuals of the $2$:$1$ solution ($\mathrm{RMS}\,{=}\,2.2$\,min) and the $3$:$1$ solution ($\mathrm{RMS}\,{=}\,9.2$\,min). The $3$:$1$ model fails to reproduce the short-timescale structure. {\it Fifth panel:} TDVs and model predictions. The $2$:$1$ model also provides a superior fit to the TDVs. KOI-142's TTV solution appears unambiguously unique.}
\label{fig:KOI142_TTVs}
\end{figure*}

\section{Reassessing the TTV planets}
\label{sec:indiv_TTV_results}

We now reassess the 12 TTV planets reported in the literature. We present them in order of decreasing evidentiary strength: from systems with compelling evidence for a unique solution, to those with plausibly unique solutions, to systems admitting multiple distinct families of solutions, and finally to cases in which the evidence for a perturbing planet is weak.

\subsection{KOI-142: The King of TTVs}
\label{sec:KOI142}

KOI-142 (Kepler-88) is a G-type star with mass $0.985~M_\odot$ and radius $0.900~R_\odot$ \citep{Fulton&Petigura2018}. The {\it Kepler} mission identified a transiting planet around KOI-142 with a period of $10.9$~days and a radius of $2.5~R_\oplus$ \citep{Borucki2011}. Soon after its discovery, the planet's extraordinarily large TTVs were recognized by \citet{Ford2011}, \citet{Steffen2012a}, and \citet{Mazeh2013}. The star's relative brightness among {\it Kepler} stars ($V\,{=}\,13$) enabled transit time measurements with uncertainties of order $1$\,min. The dominant TTV signal has a timescale of ${\approx}\,600$\,days and an amplitude of ${\approx}\,12$\,hours (Figure~\ref{fig:KOI142_TTVs}), resulting in a single-point signal-to-noise ratio of about $700$.

\citet{Nesvorny2013} modeled the TTVs and TDVs using direct $N$-body integrations with \texttt{swift\_rmvs3} \citep{Levison&Duncan1994}, recording transit times and durations. They explored parameter space using the \texttt{MultiNest} nested sampling algorithm and identified a solution involving a giant perturber of mass ${\approx}\,0.6\,M_\mathrm{J}$ and period ${\approx}\,22$\,days, placing it near the $2$:$1$ MMR. They argued that the detection of short-timescale chopping variations established the solution as unique.

Because the published study did not provide individual transit times and durations, we re-derived them from the {\it Kepler} data. The light curves were obtained from the Mikulski Archive for Space Telescopes (MAST) using the \texttt{lightkurve} package \citep{Lightkurve2018}. We used short-cadence (1-min) data when available, and otherwise long-cadence (30-min) data. The dataset contained $122$ transits across $17$ {\it Kepler} survey quarters.

For each transit, we extracted a segment of data centered on the approximate midpoint and extending 1.5 transit durations on either side. Long-term photometric trends were removed by fitting a polynomial function of time to the out-of-transit data. The polynomial degree ($1$, $2$, or $3$) was selected by minimizing the Bayesian information criterion (BIC\footnote{$\mathrm{BIC}\,{=}\,\chi^2 + k \ln(n)$, where $k$ is the number of free parameters and $n$ is the number of data points \citep{Schwarz1978}.}). The light curve was normalized by dividing by the lowest-BIC polynomial model. Flux uncertainties were estimated from the standard deviation of the out-of-transit data, and an additional jitter term was included in the likelihood to allow for excess variance. To avoid biasing the detrending process and uncertainty estimates, points more than $10\sigma$ away from the local mean were discarded.

Each transit was initially fit with a \citet{Mandel&Agol2002} model using a Nelder-Mead optimizer, holding all parameters fixed to the values reported in Table~1 of \citet{Nesvorny2013} except for the transit midpoint and jitter. The measured transit times were used to perform a second iteration of the detrending process. The resulting normalized light curves were stacked and refitted to determine consensus system parameters: $a/R_\star\,{=}\,15.46$, $R_p/R_\star\,{=}\,0.03856$, $b\,{=}\,0.7935$, $u_1\,{=}\,0.3057$, and $u_2\,{=}\,0.3318$. Finally, individual transit midpoints and durations were measured by fixing all parameters to the consensus values except for the midpoints, durations, and jitter terms. All transits were inspected visually, and we discarded transits with partial coverage or serious detrending failures.

\begin{figure*}
\centering
\includegraphics[width=0.975\textwidth]{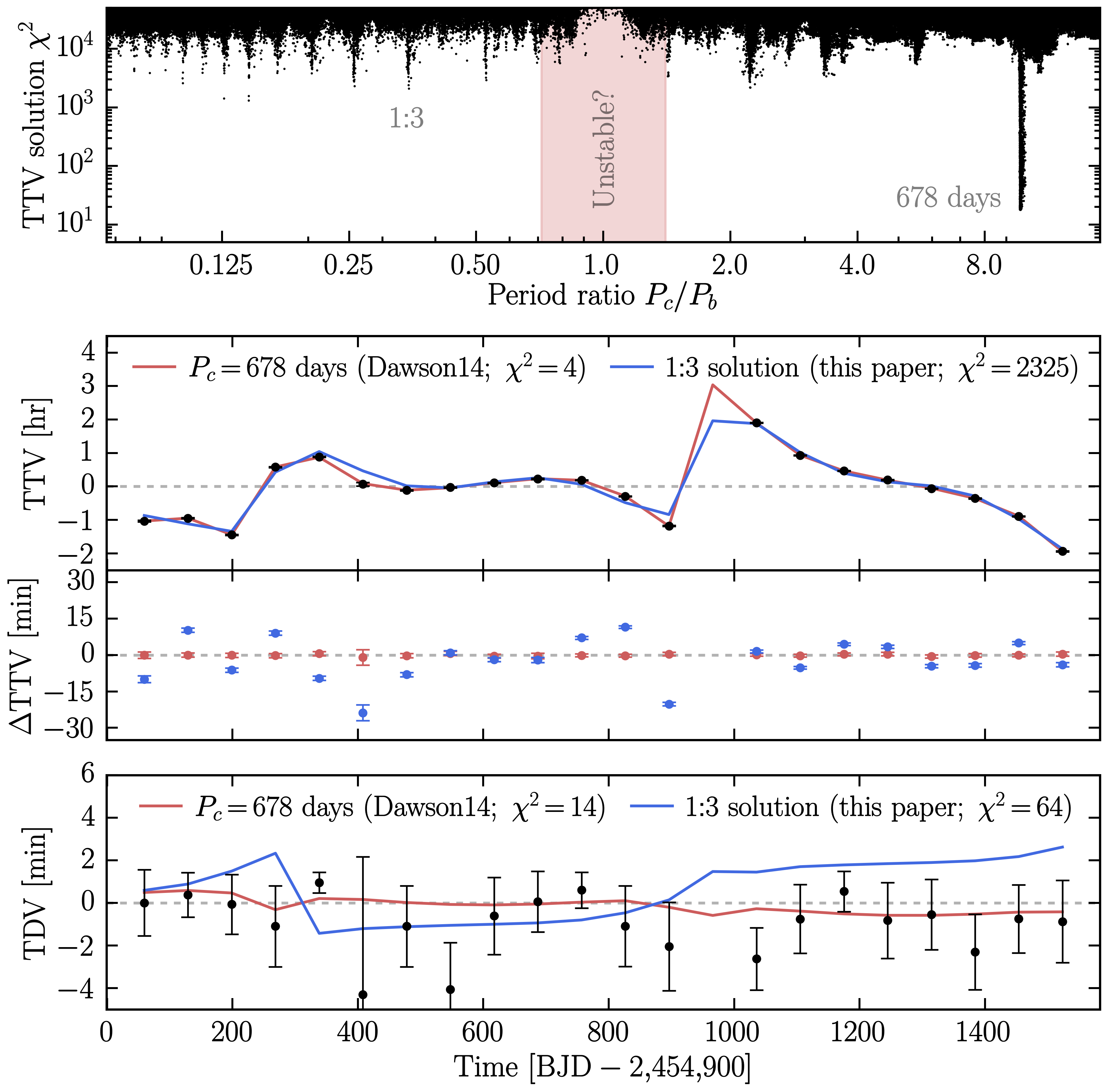}
\caption{{\bf Kepler-419}. {\it Top panel:} $\chi^2$ versus period ratio ($N_\mathrm{dof}\,{=}\,29$). The $678$-day solution is far superior to all other solutions. {\it Second panel:} TTVs measured by \citet{Dawson2014}. The red curve shows the best variant of the $P_c\,{=}\,678$\,days solution \citep{Dawson2014}. The red curve shows the $1$:$3$ solution ($P_c\,{=}\,24$\,days), which is second-best. {\it Third panel:} Residuals of the $678$-day solution ($\mathrm{RMS}\,{=}\,0.4$\,min) and the $1$:$3$ solution ($\mathrm{RMS}\,{=}\,8.9$\,min). {\it Fourth panel:} TDVs and predictions of the $678$-day and $24$-day models. Both models predict little variation, but the $678$-day model provides a better fit. It seems safe to conclude that Kepler-419's TTV solution is unique.}
\label{fig:Kepler419_TTVs}
\end{figure*}

Posterior distributions were sampled using \texttt{emcee} \citep{Foreman-Mackey2013}, which is based on the affine-invariant MCMC algorithm of \citet{Goodman&Weare2010}. We used $100$ walkers with $10^4$ steps each, discarding the first $5{,}000$ steps as burn-in. The resulting transit times and durations are shown in Figure~\ref{fig:KOI142_TTVs}.

We searched for TTV solutions using the targeted strategy described in Section~\ref{sec:TTV_modeling}. Figure~\ref{fig:KOI142_TTVs} displays $\chi^2$ as a function of the period ratio between the perturber and the transiting planet. The narrow local minima mark different solution families, associated with perturbers near different MMRs. The minimum near the $2$:$1$ MMR is by far the deepest. Table~\ref{table:KOI142_solns} lists the parameters.\footnote{We report many decimal places to facilitate reproducibility, given the extreme sensitivity of TTVs to orbital parameters.} The best-fitting perturber has a mass of $0.67~M_\mathrm{J}$, a period of $22$~days, and an eccentricity below $0.1$, in agreement with \citet{Nesvorny2013}. The total $\chi^2$ is $400.4$ for $231$ degrees of freedom ($244\,{-}\,13$). Most of the excess $\chi^2$ arises from the TDV data; the fit to the transit times alone is satisfactory (see Figure~\ref{fig:KOI142_TTVs}), suggesting that the TDV uncertainties are underestimated.

The second-best solution involves a $2.7~M_\mathrm{J}$ perturber near the $3$:$1$ MMR. Although this model reproduces the dominant $12$-hour amplitude, $600$-day signal, it fails to describe the short-timescale chopping variations and consequently provides a much poorer fit ($\chi^2\,{=}\,4554$).

To visualize the differences between the models, the third panel of Figure~\ref{fig:KOI142_TTVs} shows the TTVs after subtracting a smooth interpolating function obtained via Gaussian process (GP) regression. We used a radial basis function kernel with a minimum length scale equal to $1$\% of the observational baseline, ensuring that short-timescale variations were not absorbed by the GP. This subtraction effectively acts as a high-pass filter, isolating the chopping signal that favors the $2$:$1$ model. The structured residuals of the $3$:$1$ model are evident in the fourth panel of Figure~\ref{fig:KOI142_TTVs}. In addition, only the $2$:$1$ model reproduces the long-term TDV trend (bottom panel). 

The case for a unique solution for the perturber of KOI-142\,b is therefore exceptionally strong. This solution has been confirmed via RV observations by \citet{Barros2014}.

\begin{table}
\centering
\caption{{\bf KOI-142}. Parameters for Two Models Fit to the TTV and TDV Data}
\begin{tabular}{ccc}
 \hline
 Parameter & $2$:$1$ Solution & $3$:$1$ Solution \\
 \hline
 $m_b$ [$M_\oplus$] & $9.88323$ & $2.36489$\\ 
 $P_b$ [days] & $10.9306534$ & $10.9222381$\\ 
 $e_b \cos(\omega_b)$ & $-0.0464026$ & $-0.0280408$\\ 
 $e_b \sin(\omega_b)$ & $-0.0067883$ & $-0.0342364$\\ 
 $i_b$ [deg] & $88.96326$ & $88.69719$\\ 
 $M_b$ [deg] & $253.57129$ & $209.44999$\\ 
\hline
 $m_c$ [$M_\oplus$] & $213.19702$ & $862.97936$\\ 
 $P_c$ [days] & $22.2644586$ & $33.4046014$\\ 
 $e_c \cos(\omega_c)$ & $0.0566354$ & $0.1308937$\\ 
 $e_c \sin(\omega_c)$ & $-0.0004628$ & $0.0960239$\\ 
 $i_c$ [deg] & $85.88928$ & $77.06310$\\ 
 $\Omega_c$ [deg] & $0.29230$ & $-3.21563$\\ 
 $M_c$ [deg] & $172.05317$ & $103.05229$\\ 
\hline
 $\rho_b$ [g cm$^{-3}$] & $1.0$ & $0.2$\\ 
 $i_\mathrm{mut}$ [deg] & $3.1$ & $12.1$\\ 
\hline
$\chi^2$ & $400.4$ & $4554.0$\\ 
$N_\mathrm{dof}$ & $231$ & $231$\\
\hline
\end{tabular}
\tablecomments{The $2$:$1$ solution was reported by \citet{Nesvorny2013}, and the $3$:$1$ solution was the best alternative we found.}
\label{table:KOI142_solns}
\end{table}

\subsection{Kepler-419: The Maverick of TTVs}
\label{sec:Kepler419}

Kepler-419 (KOI-1474) is an F-type star with mass $1.427~M_\odot$ and radius $1.819~R_\odot$ \citep{Fulton&Petigura2018} accompanied by a transiting planet of radius $11~R_\oplus$ and period $70$~days \citep{Borucki2011}. Similarly to KOI-142, the star's relative brightness ($V\,{=}\,13$) allows for timing uncertainties of order $1$~min.

\citet{Dawson2012} inferred from the transit duration and period that the orbit of Kepler-419\,b is highly eccentric ($e\,{\approx}\,0.8$), a conclusion later confirmed by RV observations. They also detected multi-hour TTVs (Figure~\ref{fig:Kepler419_TTVs}). Through dynamical modeling, \citet{Dawson2014} showed that the TTVs are best explained by a ${\approx}\,7~M_\mathrm{J}$ planet on a ${\approx}\,675$-day orbit. Unlike most known TTV systems, the perturber is not near a strong MMR. Instead, the TTV signal arises primarily from the time-varying gravitational force as the eccentric transiting planet approaches and recedes from the outer companion. \citet{Dawson2014} argued that this distinct physical mechanism is what enabled a unique solution.

For our analysis, we used the transit times listed in the \texttt{TAP} column in Table~5 of \citet{Dawson2014}. We extracted impact parameter measurements and their uncertainties from Table~6 of that work, and converted them to transit durations and uncertainties. The TTV search was conducted as described in Section~\ref{sec:TTV_modeling}, with a modification: because resonance was not expected to play a dominant role in this system, we drew the period ratio $\mathcal{P}$ uniformly from $\mathcal{U}(1.15,~15)$ rather than concentrating the search on MMRs.

Figure~\ref{fig:Kepler419_TTVs} shows $\chi^2$ as a function of $\mathcal{P}$. Solutions with $P_c\,{\approx}\,678$\,days form a sharply defined family with a minimum $\chi^2$ several orders of magnitude lower than that of all other solution families. The best-fitting member of the $678$~day family (Table~\ref{table:Kepler419_solns}) has an eccentric transiting planet ($e\,{\approx}\,0.8$) and a $7.7~M_\mathrm{Jup}$ perturber on a mildly eccentric orbit, in agreement with \citet{Dawson2014}. This fit yields $\chi^2\,{=}\,17.7$ for $29$ degrees of freedom ($42\,{-}\,13$). In this solution the mutual inclination is $9^\circ$, although other members of the same solution family with $i_\mathrm{mut}\,{<}\,3^\circ$ provide statistically indistinguishable fits ($\Delta \chi^2\,{<}\,2$).

\begin{figure*}
\centering
\includegraphics[width=0.95\textwidth]{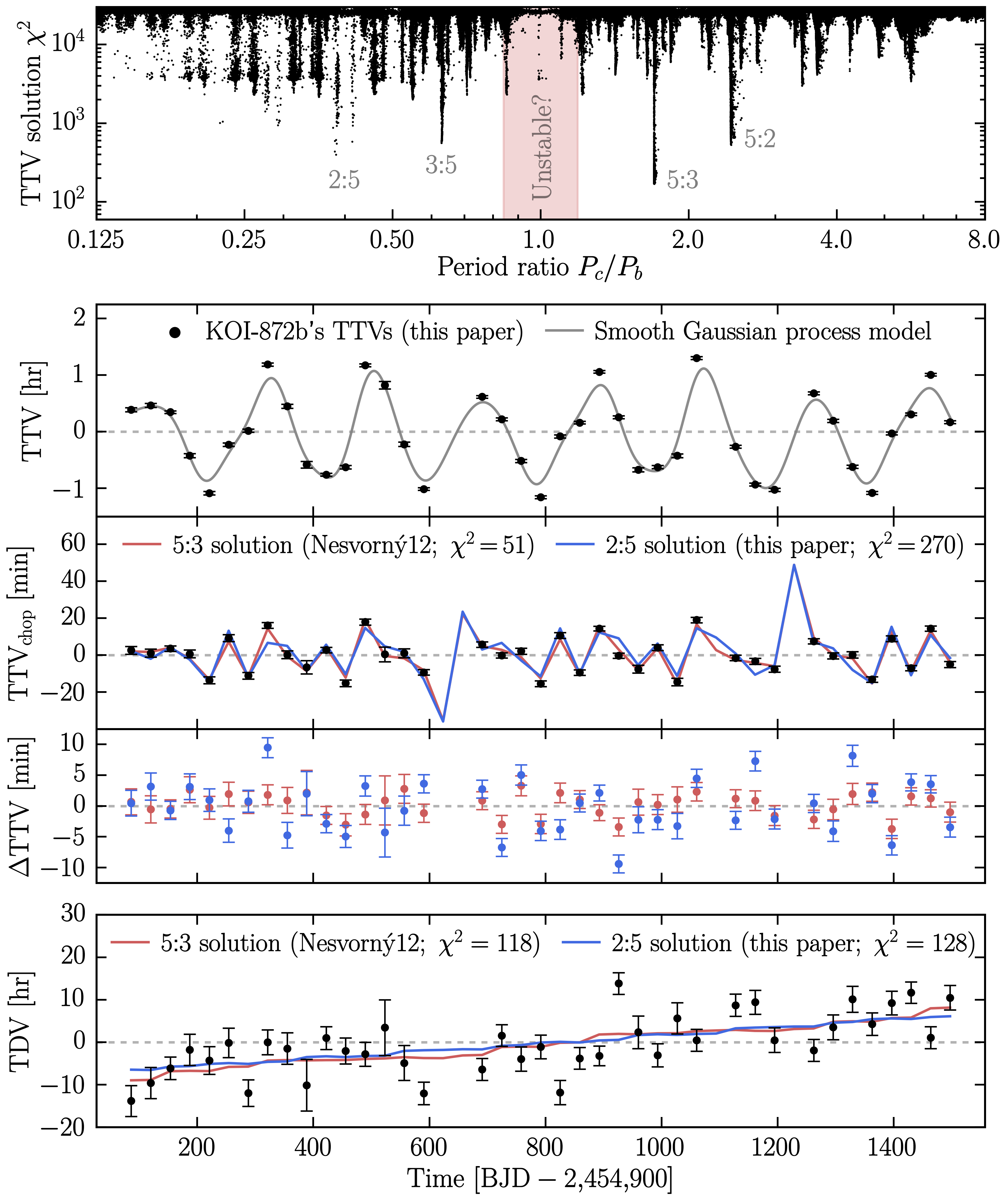}
\caption{{\bf KOI-872}. {\it Top panel:} $\chi^2$ versus period ratio ($N_\mathrm{dof}\,{=}\,65$). The $5$:$3$ solution provides the best fit, although the $2$:$5$, $5$:$2$, and $3$:$5$ solutions are also good. {\it Second panel:} TTVs from this work (Section~\ref{sec:KOI142}). The gray curve is a GP model, used as a smooth interpolant to separate long- and short-timescale variations. {\it Third panel:} High-pass-filtered TTVs obtained by subtracting the GP. The red curve shows the best variant of the $5$:$3$ solution \citep{Nesvorny2012}; the blue curve shows the $2$:$5$ solution, which is second-best. {\it Fourth panel:} Residuals of the $5$:$3$ solution ($\mathrm{RMS}\,{=}\,1.8$\,min) and the $2$:$5$ solution ($\mathrm{RMS}\,{=}\,4.2$\,min). {\it Fifth panel:} TDVs and predictions of the $5$:$3$ and $2$:$5$ models. By eye, it is difficult to declare a winner, although quantitatively, KOI-872's TTV solution does appear unique.}
\label{fig:KOI872_TTVs}
\end{figure*}

The next-best solutions involve perturbers near strong MMRs. The best $1$:$3$ solution is listed in Table~\ref{table:Kepler419_solns}, and its TTV and TDV predictions are shown in Figure~\ref{fig:Kepler419_TTVs}. Although this model reproduces the qualitative TTV morphology, it fails to match the data at the level required by the minute-scale timing uncertainties, resulting in a $\chi^2$ more than $2000$ units larger than that of the $678$-day solution. Furthermore, the $1$:$3$ model requires a substantially lower eccentricity for Kepler-419\,b than permitted by the RV data and transit shape. 

We found no alternative solution families of comparable quality to the $678$-day family. We therefore reached the same conclusion as \citet{Dawson2014} that the TTV solution for Kepler-419 is unique. RV follow up by \citet{Almenara2018} has confirmed this solution.

\subsection{KOI-872: The Prince of TTVs}
\label{sec:KOI872}

\begin{table}
\centering
\caption{{\bf Kepler-419}. Parameters for Two Models Fit to the TTV and TDV Data}
\begin{tabular}{ccc}
 \hline
 Parameter & $678$~day Solution & $1$:$3$ Solution\\
 \hline
 $m_b$ [$M_\oplus$] & $1033.21498$ & $222.72809$\\ 
 $P_b$ [days] & $69.7257768$ & $69.7245780$\\ 
 $e_b \cos(\omega_b)$ & $0.0686370$ & $-0.415035$\\ 
 $e_b \sin(\omega_b)$ & $0.7956522$ & $0.2222036$\\ 
 $i_b$ [deg] & $86.85822$ & $88.13585$\\ 
 $M_b$ [deg] & $54.09387$ & $31.31312$\\ 
\hline
 $m_c$ [$M_\oplus$] & $2445.30829$ & $12.04847$\\ 
 $P_c$ [days] & $677.5849539$ & $24.1187213$\\ 
 $e_c \cos(\omega_c)$ & $-0.0115945$ & $-0.0412600$\\ 
 $e_c \sin(\omega_c)$ & $-0.1859876$ & $-0.0099547$\\ 
 $i_c$ [deg] & $87.42536$ & $102.20751$\\ 
 $\Omega_c$ [deg] & $8.79050$ & $-4.22616$\\ 
 $M_c$ [deg] & $219.59093$ & $190.65923$\\ 
\hline
 $\rho_b$ [g cm$^{-3}$] & $3.0$ & $0.6$\\ 
 $i_\mathrm{mut}$ [deg] & $8.8$ & $14.7$\\ 
\hline
$\chi^2$ & $17.7$ & $2388.8$\\
$N_\mathrm{dof}$ & $29$ & $29$\\
\hline
\end{tabular}
\label{table:Kepler419_solns}
\tablecomments{The $678$~day solution was reported by \citet{Dawson2014}, and the $1$:$3$ solution was the best alternative we found.}
\end{table}

KOI-872 (Kepler-46) is a relatively faint ($V\,{=}\,15$) K-type star with mass $0.883\,M_\odot$ and radius $0.797\,M_\odot$ \citep{Fulton&Petigura2018}, and a transiting planet with period $36$~days and radius $0.7~R_\mathrm{J}$ \citep{Borucki2011}. \citet{Nesvorny2012} reported large ($1$~hour) nonsinusoidal TTVs with a dominant period of about $200$~days. Using $N$-body integrations coupled with the \texttt{MultiNest} nested sampling algorithm, they concluded that the TTVs are best explained by a ${\approx}\,0.4\,M_\mathrm{J}$ planet on a $57$-day orbit, near the $5$:$3$ MMR with KOI-872\,b. That solution yielded $\chi^2\,{=}\,3.4$ for $8$ degrees of freedom. A second solution near the $5$:$2$ MMR provided a slightly worse fit to the TTVs ($\Delta \chi^2\,{=}\,17$), and was rejected because it predicted large TDVs exceeding the observed level of ${\lesssim}\,20$\,min. KOI-872\,c was therefore recognized as the first planet discovered by TTVs for which the solution was unique.

Because the individual transit times and durations were not published, we re-derived them using the procedure described in Section~\ref{sec:KOI142}. The stacked light curve is best fit with $a/R_\star\,{=}\,49.83$, $R_p/R_\star\,{=}\,0.08158$, $b\,{=}\,0.6499$, $u_1\,{=}\,0.3733$, and $u_2\,{=}\,0.5106$. From the 39 transits observed by {\it Kepler}, we measured the TTVs and TDVs shown in Figure~\ref{fig:KOI872_TTVs}.

We searched for dynamical solutions as described in Section~\ref{sec:TTV_modeling}. The top panel of Figure~\ref{fig:KOI872_TTVs} shows $\chi^2$ as a function of period ratio. The horizontal structures in this plot come from the tendency for different solution families to capture similar components of the TTV data. The best-fitting solution involves a $0.34~M_\mathrm{J}$ perturber with $P_c\,{=}\,57.2$~days (Table~\ref{table:KOI872_solns}), in good agreement with \citet{Nesvorny2012}. This model yields $\chi^2\,{=}\,169$ for $65$ degrees of freedom ($78\,{-}\,13$). Most of the excess $\chi^2$ comes from the transit duration measurements; the fit to the transit times is mostly satisfactory. This suggests that the TDV uncertainties may be underestimated.

In addition to the $5$:$3$ family of solutions, we identified a small number of alternative families that qualitatively reproduce the TTV and TDV patterns. For illustration, Figure~\ref{fig:KOI872_TTVs} shows predictions of the second-best solution, corresponding to a $0.7~M_\mathrm{J}$ companion on a $13$-day orbit near the $2$:$5$ resonance. This solution was not described by \citet{Nesvorny2012}, nor in the subsequent reanalysis of all the {\it Kepler} data by \citet{SaadOlivera2017}. The $5$:$3$ and $2$:$5$ models produce similar short-timescale TTV structure, as illustrated in the third panel of Figure~\ref{fig:KOI872_TTVs}, where long-timescale TTV trends have been removed. Despite the visual impression that both solutions are good, the $5$:$3$ model improves the fit by $\Delta \chi^2\,{\approx}\,200$ relative to the $2$:$5$ model, indicating a decisive statistical preference.

To test the reliability of the $5$:$3$ and $2$:$5$ solutions, we carried out two additional experiments. First, we performed leave-one-out cross-validation (LOO-CV), in which individual transit times and durations were removed from the fit, and following reoptimization, models were judged on their ability to correctly predict the withheld data. The $5$:$3$ model made more accurate predictions than the $2$:$5$ model, with a difference in the expected log predictive density, $\Delta\,\mathrm{elpd}$, of $65$ units (see, e.g., \citealt{Welbanks2023}). Uncertainties on the transit time predictions were estimated using the inverted Hessian matrix, and were larger for the $2$:$5$ model. Second, we evaluated the influence of outliers by comparing the solutions using an outlier-tolerant likelihood function, in which the residuals follow a Student's-t distribution. Following \citet{JontofHutter2016}, we fixed $N_\mathrm{dof}\,{=}\,2$ and determined the location and scale of the distribution by fitting the residuals of the $5$:$3$ solution. We found that the $5$:$3$ solution was still favored, with $\Delta \ln \mathcal{L}\,{=}\,47$, even when downweighting the outliers.

Our analysis suggests that KOI-872's TTV solution is unique. However, the case for a unique solution is less secure than the two previous cases, because the discrepancies are subtle and limited to $5$\,--\,$10$ data points. The best way to obtain a firm conclusion is probably RV follow-up observations. The $5$:$3$ solution predicts an RV signal with semi-amplitude $20$~m/s and period $57$~days, whereas the $2$:$5$ solution predicts a $70$~m/s RV signal with a period of $13$~days.

\begin{figure*}
\centering
\includegraphics[width=0.95\textwidth]{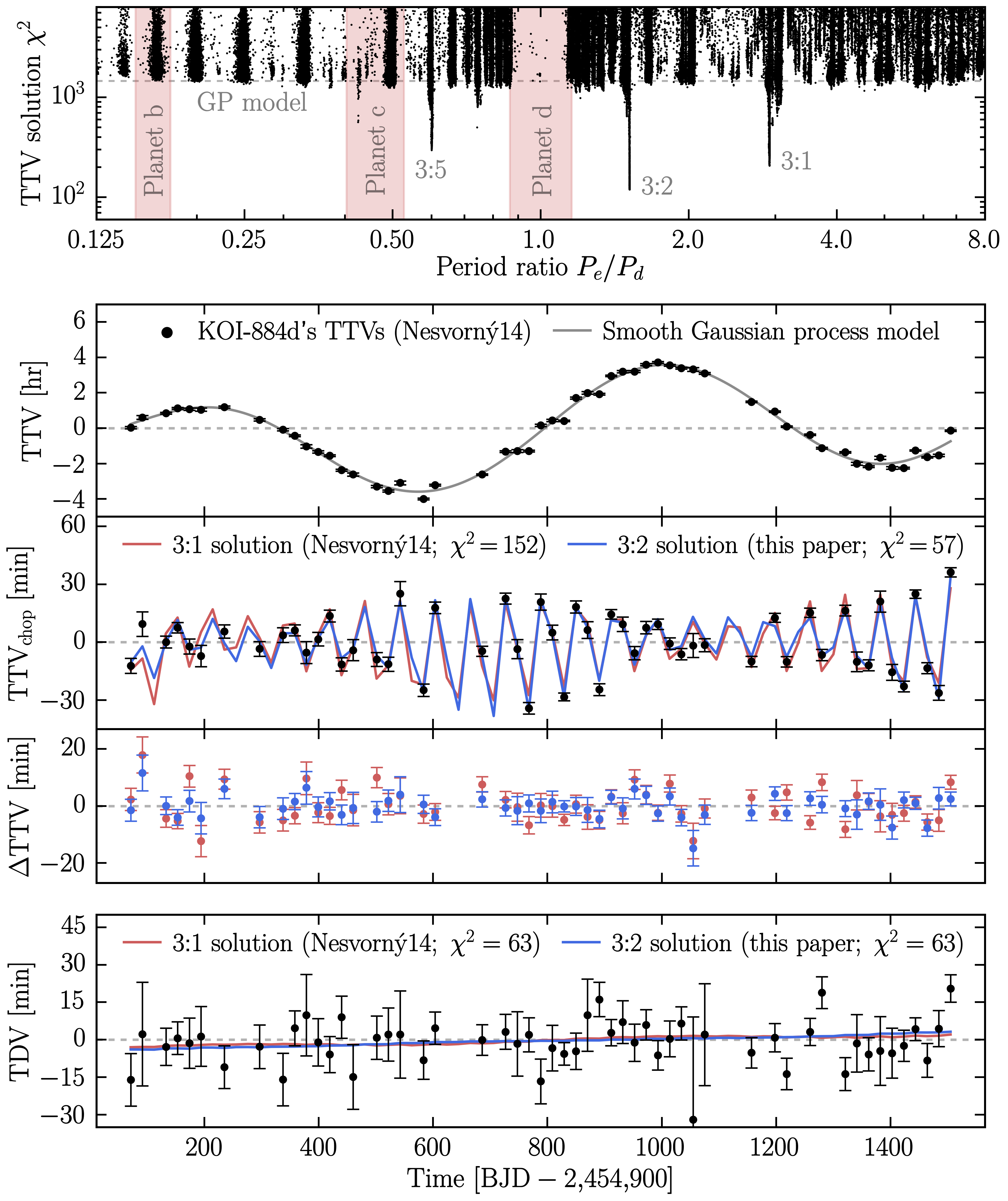}
\caption{{\bf KOI-884}. {\it Top panel:} $\chi^2$ versus period ratio ($N_\mathrm{dof}\,{=}\,95$). The $3$:$2$ solution provides the best fit, followed closely by the $3$:$1$ and $3$:$5$ solutions. {\it Second panel:} TTVs from \citet{Nesvorny2014}, along with a GP model that provides a smooth interpolation to separate long- and short-timescale variations. The $\chi^2$ value associated with the GP model is marked by a horizontal gray line in the top panel; many solutions can reproduce this smooth variation. {\it Third panel:} High-pass-filtered TTVs obtained by subtracting the GP. The red curve shows the best variant of the $3$:$1$ solution \citep{Nesvorny2013}. The blue curve shows the $3$:$2$ solution, which provides a somewhat superior fit. {\it Fourth panel:} Residuals of the $3$:$2$ solution ($\mathrm{RMS}\,{=}\,4.0$\,min) and the $3$:$1$ solution ($\mathrm{RMS}\,{=}\,6.0$\,min) {\it Fifth panel:} TDVs and predictions of the $3$:$2$ and $3$:$1$ models. KOI-884b's TTV solution is at least borderline unique, but the best solution has not been reported in the literature.}
\label{fig:KOI884_TTVs}
\end{figure*}

\subsection{KOI-884: The Trickster of TTVs}
\label{sec:KOI884}

KOI-884 (Kepler-247) is a relatively faint ($V\,{=}\,15$) K-type star with mass $0.887\,M_\odot$ and radius $0.788\,R_\odot$ \citep{Fulton&Petigura2018}. Three transiting planets are known, with orbital periods of $3.3$, $9.4$, and $20.5$~days \citep{Borucki2011}. Planet b has a radius of $1.6~R_\oplus$. Planets c and d are larger, with radii of $4.3~R_\mathrm{\oplus}$ and $5.0~R_\mathrm{\oplus}$ \citep{Fulton&Petigura2018}.

\begin{table}
\centering
\caption{{\bf KOI-872}. Parameters for Two Models Fit to the TTV and TDV Data}
\begin{tabular}{ccc}
 \hline
 Parameter & $5$:$3$ Solution & $2$:$5$ Solution \\
 \hline
 $m_b$ [$M_\oplus$] & $132.57577$ & $96.48064$\\ 
 $P_b$ [days] & $33.5692945$ & $33.7399643$\\ 
 $e_b \cos(\omega_b)$ & $0.0216773$ & $-0.2382234$\\ 
 $e_b \sin(\omega_b)$ & $-0.0154956$ & $0.0664965$\\ 
 $i_b$ [deg] & $89.30509$ & $89.37997$\\ 
 $M_b$ [deg] & $281.07029$ & $109.35935$\\ 
\hline
 $m_c$ [$M_\oplus$] & $109.59135$ & $226.90154$\\ 
 $P_c$ [days] & $57.1897813$ & $12.9537653$\\ 
 $e_c \cos(\omega_c)$ & $0.0239285$ & $-0.1429947$\\ 
 $e_c \sin(\omega_c)$ & $0.0004911$ & $0.0325668$\\ 
 $i_c$ [deg] & $90.38059$ & $92.23877$\\ 
 $\Omega_c$ [deg] & $-0.65193$ & $-0.79512$\\ 
 $M_c$ [deg] & $179.68662$ & $98.82002$\\ 
\hline
 $\rho_b$ [g cm$^{-3}$] & $2.0$ & $1.5$\\ 
 $i_\mathrm{mut}$ [deg] & $1.3$ & $3.0$\\ 
\hline
$\chi^2$ & $169.1$ & $397.7$\\ 
$N_\mathrm{dof}$ & $65$ & $65$\\ 
\hline
\end{tabular}
\label{table:KOI872_solns}
\tablecomments{The $5$:$3$ solution was reported by \citet{Nesvorny2012}, and the $2$:$5$ solution was the best alternative we found.}
\end{table}

\citet{Mazeh2013} reported TTVs for planet d with a period of ${\approx}\,800$\,days and an amplitude of ${\approx}\,4$~hours (Figure~\ref{fig:KOI884_TTVs}). Using $N$-body integrations and \texttt{multinest}, \citet{Nesvorny2014} concluded that the TTVs and TDVs implicate a perturber of mass ${\approx}\,2.6\,M_\mathrm{J}$ on a $60$-day orbit, near the $3$:$1$ MMR.\footnote{Although a unique family of masses and periods was reported, a discrete degeneracy still remained between solutions with mutual inclinations of $1^\circ$ and $14^\circ$.} They attributed the uniqueness of the solution to the high TTV signal-to-noise ratio (about $55$), broad temporal coverage (nearly twice $P_\mathrm{TTV}$), and clear detection of short-timescale chopping signals.

We adopted transit times and durations from Table~2 of \citet{Nesvorny2014} and performed our usual search for two-planet TTV solutions (Section~\ref{sec:TTV_modeling}). The top panel of Figure~\ref{fig:KOI884_TTVs} shows $\chi^2$ versus period ratio. The plateau near $\chi^2\,{\approx}\,1000$ is formed by solutions that can reproduce the smooth, long-timescale component of the TTV signal but not the chopping signal.

Our search recovered the $3$:$1$ solution family, for which the best-fitting members imply a mass of ${\approx}\,1000\,M_\oplus$ for planet d, corresponding to an implausibly large bulk density of about $40$\,g\,cm$^{-3}$. However, closely related variants within the same family yield $m_d\,{\approx}\,30\,M_\oplus$ (Table~\ref{table:KOI884_solns}), with only a marginally worse fit ($\Delta \chi^2\,{\approx}\,8$). The derived masses, eccentricities, mutual inclinations, and $\chi^2$ values are broadly consistent with those reported by \citet{Nesvorny2014}. The parameters of the coplanar $3$:$1$ solution and the predicted TTVs and TDVs are shown in Figure~\ref{fig:KOI884_TTVs}.

In addition to the $3$:$1$ solution, our search uncovered a distinct $3$:$2$ solution in which the perturber has an orbital period of $31$~days and a mass of ${\approx}\,66\,M_\oplus$. This model provides a statistically superior fit to the data, with $\chi^2\,{=}\,119$ for $95$ degrees of freedom ($108\,{-}\,13$), improving on the best $3$:$1$ solution by $\Delta \chi^2\,{=}\,95$. The residuals exhibit little structure, and the improvement stems primarily from a more accurate reproduction of the short-timescale TTV features (Figure~\ref{fig:KOI884_TTVs}, third panel).

The $3$:$2$ solution is also physically reasonable. It assigns a mass of ${\approx}\,20\,M_\oplus$ to planet d, implying a bulk density of $0.8$\,g\,cm$^{-3}$, typical for planets of radius $5\,R_\oplus$. The nontransiting companion has a mass of ${\approx}\,66~M_\oplus$, more representative of close-in planets than the $2.6~M_\mathrm{J}$ companion required by the $3$:$1$ solution. Despite broad priors on eccentricities and inclinations, the $3$:$2$ solution naturally converges toward low eccentricities (${\lesssim}\,0.1$) and a low mutual inclination (${\approx}\,1^\circ$). The existence of this $3$:$2$ family is notable because \citet{Nesvorny2014} reported that earlier analyses of quarters $0$\,--\,$12$ suggested such a possibility, but also that including quarters $13$\,--\,$17$ significantly worsened the fit ($\chi^2\,{>}\,1000$). In contrast, our $3$:$2$ solution fits the full datasets (quarters $0$\,--\,$17$) comparably well across all epochs.

Satisfactory solutions must also avoid predicting TTVs for KOI-884\,c in excess of ${\approx}\,5$\,min. As a consistency check, we measured planet c's TTVs and TDVs using the process described for KOI-142 in Section~\ref{sec:KOI142}, and we carried out a three-planet TTV solution search. The results were similar to the two-planet case. Again, the $3$:$2$ and $3$:$1$ solutions emerged as most promising, with similar parameters to those in Table~\ref{table:KOI884_solns}, except $m_d$ was somewhat smaller. The preference for the $3$:$1$ solution also increased ($\Delta \chi^2\,{\approx}\,150$).

Because the statistical preference for the $3$:$2$ solution is significant but not overwhelming, its status depends critically on the accuracy of the timing uncertainties. To test robustness, we re-derived the transit times and durations for KOI-884\,d using the process described in Section~\ref{sec:KOI142}. With these independently measured TTVs and TDVs, the $3$:$2$ solution remains favored over the $3$:$1$ solution by $\Delta \chi^2\,{\approx}\,100$.

As we did for KOI-872, we performed two additional tests between KOI-884's competing TTV solutions: LOO-CV and a heavy-tailed likelihood function. Both tests ruled in favor of the $3$:$2$ solution; this solution predicts holdout times more accurately ($\Delta\,\mathrm{elpd}\,{=}\,59$) and remained the statistical favorite ($\Delta \ln \mathcal{L}\,{=}\,21$) when assuming the residuals obey a Student's-t instead of the usual Gaussian distribution.

We therefore consider the TTV solution for KOI-884 to have narrowly cleared the threshold for uniqueness, with a configuration that differs from the originally reported $3$:$1$ solution. The $3$:$1$ solution predicts a much larger RV signal ($140$~m/s semiamplitude) than the $3$:$2$ solution ($15$~m/s semiamplitude).

\begin{figure*}
\centering
\includegraphics[width=0.95\textwidth]{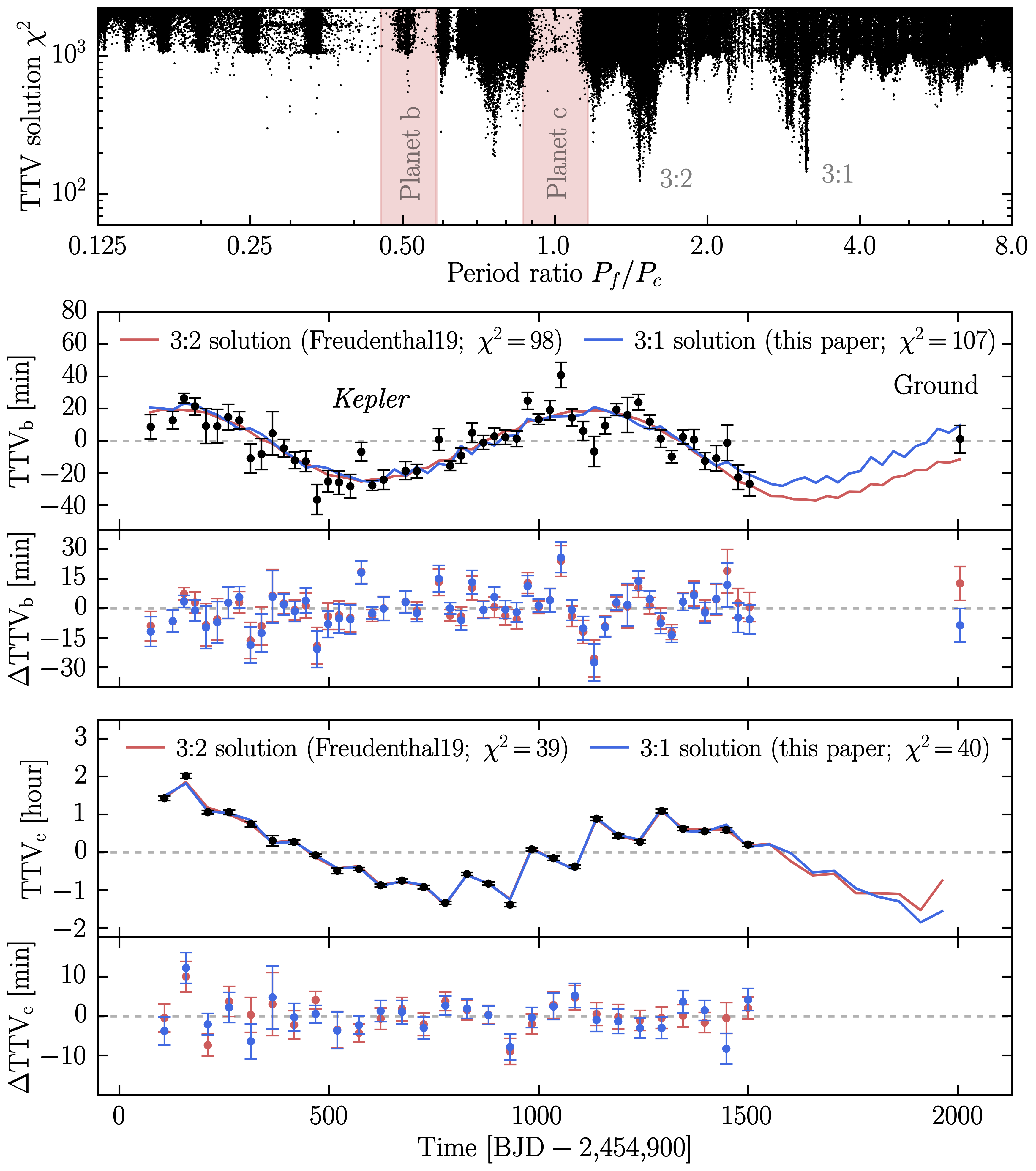}
\caption{{\bf Kepler-82}. {\it Top panel:} $\chi^2$ versus period ratio ($N_\mathrm{dof}\,{=}\,61$). The $3$:$2$ and $3$:$1$ solutions provide the best fit to the data. {\it Second panel:} Planet b's TTVs from {\it Kepler} and ground-based follow up, as measured in this work. The red curve shows the best variant of the $3$:$2$ solution \citep{Freudenthal2019}, and the blue curve shows the $3$:$1$ solution. {\it Third panel:} Planet b's residuals for the $3$:$2$ solution ($\mathrm{RMS}\,{=}\,9.1$\,min) and the $3$:$1$ solution ($\mathrm{RMS}\,{=}\,9.6$\,min). {\it Fourth panel:} Planet c's TTVs, as measured from this work, along with the $3$:$2$ and $3$:$1$ models. {\it Fifth panel:} Planet c's residuals for the $3$:$2$ solution ($\mathrm{RMS}\,{=}\,3.7$\,min) and the $3$:$1$ solution ($\mathrm{RMS}\,{=}\,4.2$\,min). The $3$:$2$ solution provides a minor improvement, with $\Delta \chi^2\,{=}\,10$, but the timing uncertainties are likely to be underestimated. The differences are too slim to conclusively identify the correct solution.}
\label{fig:Kepler82_TTVs}
\end{figure*}

\begin{table}
\centering
\caption{{\bf KOI-884}. Parameters for Two Models Fit to the TTV and TDV Data}
\begin{tabular}{ccc}
 \hline
 Parameter & $3$:$1$ Solution & $3$:$2$ Solution \\
 \hline
 $m_d$ [$M_\oplus$] & $32.65949$ & $18.90145$\\ 
 $P_d$ [days] & $20.5349819$ & $20.5013918$\\ 
 $e_d \cos(\omega_d)$ & $-0.0944342$ & $-0.0292275$\\ 
 $e_d \sin(\omega_d)$ & $0.0001264$ & $0.0092450$\\ 
 $i_d$ [deg] & $88.93142$ & $88.93399$\\ 
 $M_d$ [deg] & $111.97909$ & $120.94706$\\ 
\hline
 $m_e$ [$M_\oplus$] & $806.52811$ & $66.05405$\\ 
 $P_e$ [days] & $59.9053798$ & $31.1016590$\\ 
 $e_e \cos(\omega_e)$ & $-0.0340395$ & $-0.0183267$\\ 
 $e_e \sin(\omega_e)$ & $0.0229054$ & $0.0131596$\\ 
 $i_e$ [deg] & $88.51850$ & $88.73887$\\ 
 $\Omega_e$ [deg] & $-0.24293$ & $-0.33910$\\ 
 $M_e$ [deg] & $171.36592$ & $112.36320$\\ 
\hline
 $\rho_d$ [g cm$^{-3}$] & $1.5$ & $0.8$\\ 
 $i_\mathrm{mut}$ [deg] & $0.5$ & $0.4$\\ 
\hline
$\chi^2$ & $214.6$ & $119.2$\\ 
$N_\mathrm{dof}$ & $95$ & $95$\\ 
\hline
\end{tabular}
\label{table:KOI884_solns}
\tablecomments{The $3$:$1$ solution was reported by \citet{Nesvorny2012}, and we found the superior $3$:$2$ solution.}
\end{table}

\subsection{Kepler-82: The Janus of TTVs}
\label{sec:Kepler82}

Kepler-82 (KOI-880) is a G-type star with $V\,{=}\,15$, mass $0.946~M_\odot$, and radius $0.835~R_\odot$ \citep{Fulton&Petigura2018}. Four transiting planets have been detected, with periods $2.4$, $5.9$, $26.4$, and $51.5$~days \citep{Borucki2011}. From inside out, their radii are $1.6$, $2.4$, $4.2$, and $5.5~R_\oplus$ \citep{Fulton&Petigura2018}. The outer two planets (b and c) were confirmed by virtue of large, approximately sinusoidal TTVs by \citet{Xie2013}, and the inner two planets (d and e) were statistically validated by \citet{Rowe2014}. The TTVs of planets b and d were used by \citet{Xie2013} and \citet{Hadden&Lithwick2014} to estimate their masses. Later, \citet{Ofir2018} noticed short-timescale structure in planet c's TTVs that could not be explained solely by the interaction between b and c. \citet{Freudenthal2019} performed dynamical modeling including three ground-based transit observations, and proposed the existence of an additional $20~M_\oplus$ perturber on a $76$-day orbit.

Because individual transit times were not provided by \citet{Freudenthal2019}, we re-derived the transit times for Kepler-82\,b and c using the procedure described in Section~\ref{sec:KOI142}. The stacked light curve for planet b is best fit with $a/R_\star\,{=}\,46.35$, $R_p/R_\star\,{=}\,0.0401$, $b\,{=}\,0.5287$, $u_1\,{=}\,0.5616$, and $u_2\,{=}\,0.0000$. For planet c, we measured $a/R_\star\,{=}\,58.96$, $R_p/R_\star\,{=}\,0.0555$, $b\,{=}\,0.4162$, $u_1\,{=}\,0.5027$, and $u_2\,{=}\,0.1094$.

We also analyzed the publicly available ground-based photometry from the KOINet campaign. For each dataset, we performed a joint fit including a transit model (with parameters fixed to the {\it Kepler}-derived values), a polynomial detrending function, and a flux jitter term. For Kepler-82\,b, a linear detrending model was sufficient to recover the transit reported by \citet{Freudenthal2019}, and we estimated a timing uncertainty of $8.6$~min. For the two Kepler-82\,c datasets, however, the results were inconclusive. For the first time series, both a transit model and a flat light curve provided statistically comparable fits ($\Delta \chi^2\,{<}\,2$), and the posterior distribution was consistent with a nondetection. For the second time series, neither linear nor quadratic detrending yielded a clear transit signal. Because the detrending methodology of \citet{Freudenthal2019} is not described in detail, and because the signal-to-noise ratio is low, we excluded these two marginal detections from our TTV analysis. Figure~\ref{fig:Kepler82_TTVs} shows the TTVs that were retained.

We carried out the TTV solution search process as described in Section~\ref{sec:TTV_modeling}, with results shown in the top panel of Figure~\ref{fig:Kepler82_TTVs}. The formally best-fitting solutions require large mutual inclinations between planets b and c ($i_\mathrm{mut}\,{\gtrsim}\,20^\circ$), a seemingly unlikely configuration given that both planets transit. Slightly worse solutions ($\Delta \chi^2\,{\approx}\,12$) allow for more nearly coplanar configurations (${\lesssim}\,8^\circ$), which seem more plausible. In both cases, the lowest $\chi^2$ values are obtained when the additional perturber has $P_f\,{\approx}\,76$\,days, near the $3$:$2$ MMR with planet c (Table~\ref{table:Kepler82_solns}). The best nearly coplanar $3$:$2$ solution yields $\chi^2\,{=}\,137$ for $61$ degrees of freedom ($81\,{-}\,20$). The excess $\chi^2$ suggests the timing uncertainties are underestimated, particularly for planet b.

The $3$:$2$ solution is not unique. We also identified a $3$:$1$ solution involving a ${\approx}\,600\,M_\oplus$ perturber that reproduces the TTV data comparably well (\citealt{Freudenthal2019} also identified a variant of this solution family). Some members of this family have implausibly large densities for planet b, but other variants have reasonable planet densities, low eccentricities, and low mutual inclinations (Table~\ref{table:Kepler82_solns}). The best-fitting plausible $3$:$1$ solution gives $\chi^2\,{=}\,147$, only $\Delta \chi^2\,{\approx}\,10$ worse than the best $3$:$2$ solution. If the timing uncertainties are inflated so that the $3$:$2$ solution achieves $\chi^2\,{=}\,N_\mathrm{dof}$, the difference between the two models decreases to just $\Delta \chi^2\,{\approx}\,4$.

Thus, even with the nominal (and likely underestimated) uncertainties, the statistical preference for the $3$:$2$ solution is minor. Given the weakness of the ground-based transit detections and the sensitivity of the model comparison to the assumed uncertainties, we conclude that the TTV solution for Kepler-82 is not unique. There are two distinct families of viable solutions, near the $3$:$1$ and $3$:$2$ MMRs.

\begin{table}
\centering
\caption{{\bf Kepler-82}. Parameters for Two Models Fit to the TTV Data}
\begin{tabular}{ccc}
 \hline
 Parameter & $3$:$2$ Solution & $3$:$1$ Solution\\
\hline
 $m_b$ [$M_\oplus$] & $17.75698$ & $53.18625$\\ 
 $P_b$ [days] & $26.4435820$ & $26.4427517$\\ 
 $e_b \cos(\omega_b)$ & $-0.0042705$ & $-0.0087922$\\ 
 $e_b \sin(\omega_b)$ & $-0.0095463$ & $0.0016590$\\ 
 $i_b$ [deg] & $88.89653$ & $89.59194$\\ 
 $M_b$ [deg] & $273.89616$ & $350.93986$\\ 
\hline
 $m_c$ [$M_\oplus$] & $13.48031$ & $25.81584$\\ 
 $P_c$ [days] & $51.5298696$ & $51.5218614$\\ 
 $e_c \cos(\omega_c)$ & $-0.0220812$ & $-0.0260517$\\ 
 $e_c \sin(\omega_c)$ & $-0.0046072$ & $-0.0164473$\\ 
 $i_c$ [deg] & $89.09062$ & $89.51661$\\ 
 $\Omega_c$ [deg] & $8.09264$ & $7.61188$\\ 
 $M_c$ [deg] & $232.26231$ & $212.30570$\\ 
\hline
 $m_d$ [$M_\oplus$] & $23.38585$ & $610.73222$\\ 
 $P_d$ [days] & $75.6683003$ & $161.5146160$\\ 
 $e_d \cos(\omega_d)$ & $-0.0107340$ & $0.0315180$\\ 
 $e_d \sin(\omega_d)$ & $-0.0050353$ & $-0.0729013$\\ 
 $i_d$ [deg] & $92.75540$ & $84.55333$\\ 
 $\Omega_d$ [deg] & $9.51834$ & $-5.59606$\\ 
 $M_d$ [deg] & $183.37605$ & $19.98173$\\ 
\hline
 $\rho_b$ [g cm$^{-3}$] & $1.4$ & $4.3$\\ 
 $\rho_c$ [g cm$^{-3}$] & $0.5$ & $0.9$\\ 
 $i_\mathrm{mut,bc}$ [deg] & $8.1$ & $7.6$\\ 
 $i_\mathrm{mut,cd}$ [deg] & $3.9$ & $14.1$\\ 
\hline
$\chi^2$ & $136.6$ & $146.9$\\ 
$N_\mathrm{dof}$ & $61$ & $61$\\ 
\hline
\end{tabular}
\label{table:Kepler82_solns}
\tablecomments{\cite{Freudenthal2019} reported variants of both solutions but preferred the $3$:$2$ solution.}
\end{table}

\subsection{Kepler-411: The Hydra of TTVs}
\label{sec:Kepler411}

\begin{figure*}
\centering
\includegraphics[width=0.975\textwidth]{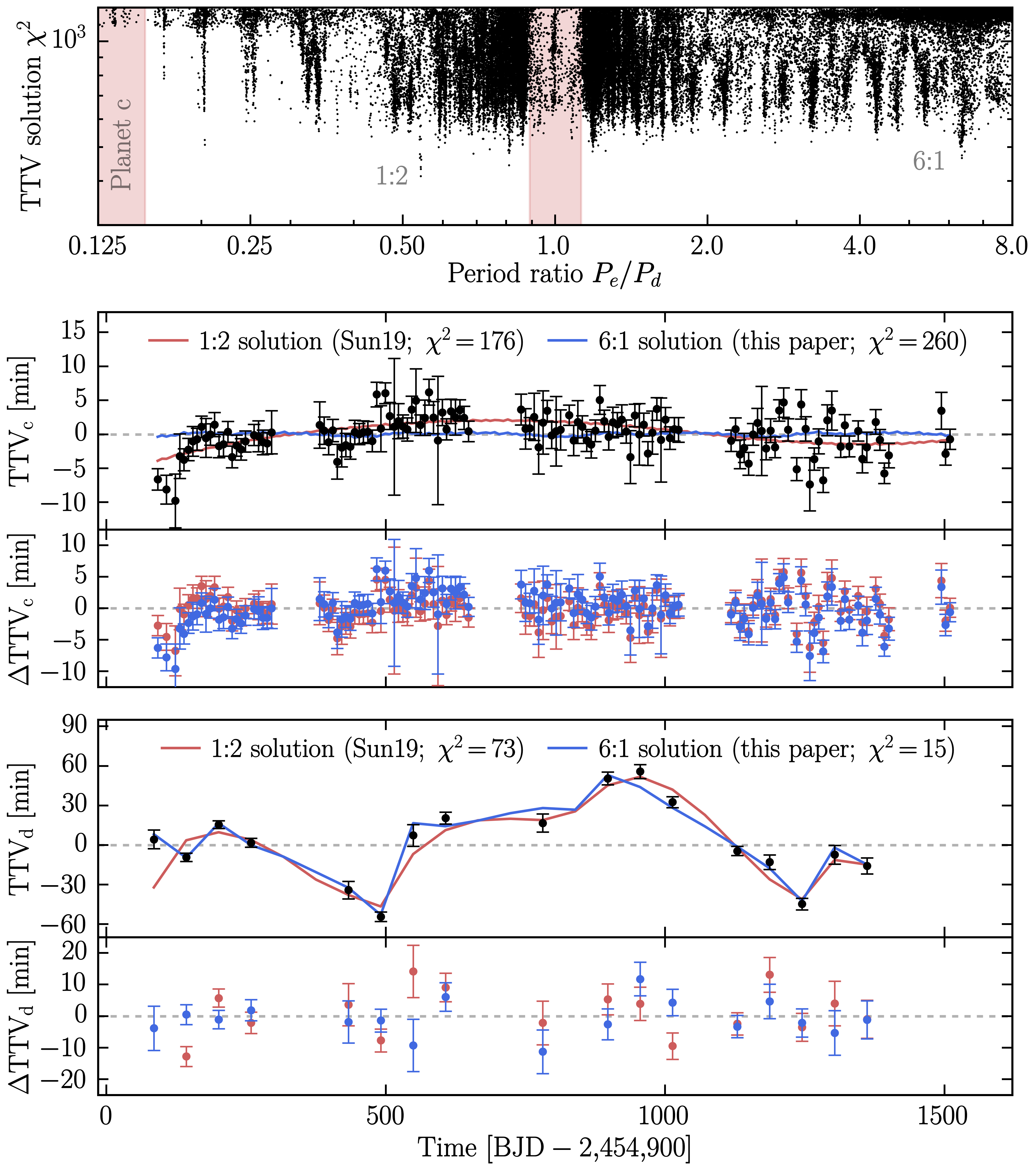}
\caption{{\bf Kepler-411}. {\it Top panel:} $\chi^2$ versus period ratio ($N_\mathrm{dof}\,{=}\,435$). Many solutions provide a good fit to the data, but most involve implausible planet densities, large eccentricities, or grossly misaligned orbits. The $1$:$2$ and $6$:$1$ solutions seem most plausible. {\it Second panel:} Planet c's TTVs from \citet{Sun2019} with inflated uncertainties (see Section~\ref{sec:Kepler411}). The red curve shows the best variant of the $1$:$2$ solution \citep{Sun2019}, and the blue curve shows the $6$:$1$ solution. {\it Third panel:} Planet c's residuals for the $1$:$2$ solution ($\mathrm{RMS}\,{=}\,2.4$\,min) and the $6$:$1$ solution ($\mathrm{RMS}\,{=}\,2.8$\,min). {\it Fourth panel:} Planet d's TTVs from \citet{Sun2019} with inflated uncertainties, along with the $1$:$2$ and $6$:$1$ models. {\it Fifth panel:} Planet d's residuals for the $1$:$2$ solution ($\mathrm{RMS}\,{=}\,11.5$\,min) and the $6$:$1$ solution ($\mathrm{RMS}\,{=}\,5.5$\,min). Planet b's flat TTVs were included in the fit but have been omitted from this figure. The $\Delta \chi^2$ formally favors the $1$:$2$ solution, but stellar activity and uncertain timing uncertainties make this preference questionable.}
\label{fig:Kepler411_TTVs}
\end{figure*}

Kepler-411 (KOI-1781) is a K-type star with mass $0.87~M_\odot$ and radius $0.82~R_\odot$ \citep{Sun2019}. It is relatively bright for the {\it Kepler} sample ($V\,{=}\,12.4$) and exhibits significant photometric variability attributed to magnetic activity \citep{Sun2019}. Transit-like signals were first identified by the Planet Hunters project \citep{Schwamb2012}, and subsequent {\it Kepler} analyses reported two short-period transiting planets ($3.0$ and $7.8$~days) and a longer-period companion at $58$~days \citep{Batalha2013, Rowe2014}. The inner two planets (b and c) were statistically validated \citep{Wang2014, Morton2016}, and TTVs of both planets were reported in early {\it Kepler} catalogs \citep{Mazeh2013, Rowe2014}. Later, \citet{Sun2019} modeled the system's TTVs, and proposed a nontransiting $10~M_\oplus$ planet on a $31.5$-day orbit between planets c and d.

We collected the transit times for Kepler-411\,b, c, and d from \citet{Sun2019} and carried out our usual TTV search process (Section~\ref{sec:TTV_modeling}). In an initial search using the published timing uncertainties, we did not find any satisfactory solutions. The best model had $\chi^2\,{=}\,1095$ for $435$ degrees of freedom ($462\,{-}\,27$). The fit to planet d's TTVs was particularly poor; the optimization tended to prioritize fitting planet c, whose reported timing uncertainties are extremely small. Given the star's substantial short-term variability, underestimated timing errors seem likely. Indeed, \citet{Sun2019} noted this concern and inflated their uncertainties prior to MCMC posterior sampling.

To address this issue, we introduced a $1.5$-min jitter term applied uniformly to all transit times. This changed the best-fit $\chi^2$ to approximately $N_\mathrm{dof}$, after which we repeated the TTV solution search. The resulting $\chi^2$ landscape as a function of period ratio is shown in the top panel of Figure~\ref{fig:Kepler411_TTVs}. We identified numerous TTV solutions of comparable statistical quality. Many, however, require implausible planet densities, large eccentricities, or strongly misaligned orbits. The extra computational cost of modeling a four-planet system limited the density of our search grid relative to the other systems considered in this work.

\begin{table}
\centering
\caption{{\bf Kepler-411}. Parameters for Three Models Fit to the TTV Data}
\begin{tabular}{cccc}
 \hline
 Parameter & $1$:$2$ Solution & $6$:$1$ Solution & $4$:$5$ Solution\\
 \hline
 $m_b$ [$M_\oplus$] & $5.67251$ & $7.12114$ & $5.96252$\\ 
 $P_b$ [days] & $3.0051658$ & $3.0053448$ & $3.0052069$\\ 
 $e_b \cos(\omega_b)$ & $-0.0028466$ & $-0.0678940$ & $0.0290601$\\ 
 $e_b \sin(\omega_b)$ & $-0.0678604$ & $-0.0291221$ & $-0.0242258$\\ 
 $i_b$ [deg] & $89.25005$ & $89.12139$ & $87.38448$\\ 
 $M_b$ [deg] & $64.74904$ & $136.78160$ & $8.47885$\\ 
 \hline
 $m_c$ [$M_\oplus$] & $14.40444$ & $42.67663$ & $45.64909$\\ 
 $P_c$ [days] & $7.8348458$ & $7.8345882$ & $7.8347187$\\ 
 $e_c \cos(\omega_c)$ & $-0.0000238$ & $0.0735771$ & $0.0223469$\\ 
 $e_c \sin(\omega_c)$ & $-0.0000183$ & $-0.0526026$ & $0.0819316$\\ 
 $i_c$ [deg] & $90.03363$ & $90.10047$ & $88.28318$\\ 
 $\Omega_c$ [deg] & $6.22601$ & $17.60421$ & $25.04299$\\ 
 $M_c$ [deg] & $337.75424$ & $221.90012$ & $118.06192$\\ 
 \hline
 $m_c$ [$M_\oplus$] & $14.32815$ & $25.69040$ & $27.79974$\\ 
 $P_c$ [days] & $58.01709$ & $58.0399057$ & $58.0356254$\\ 
 $e_c \cos(\omega_c)$ & $-0.0116594$ & $0.0053621$ & $-0.1356644$\\ 
 $e_c \sin(\omega_c)$ & $-0.1405647$ & $0.0084217$ & $0.0003624$\\ 
 $i_c$ [deg] & $89.65543$ & $90.71708$ & $89.69769$\\ 
 $\Omega_c$ [deg] & $13.76844$ & $13.00420$ & $17.13755$\\ 
 $M_c$ [deg] & $19.77908$ & $225.31388$ & $119.24652$\\ 
 \hline
 $m_c$ [$M_\oplus$] & $6.44202$ & $2736.71816$ & $7.25512$\\ 
 $P_c$ [days] & $31.3547069$ & $369.0322451$ & $47.1449666$\\ 
 $e_c \cos(\omega_c)$ & $-0.2534169$ & $0.0075567$ & $-0.1271890$\\ 
 $e_c \sin(\omega_c)$ & $-0.1151369$ & $-0.2218099$ & $0.0307309$\\ 
 $i_c$ [deg] & $69.71448$ & $64.95722$ & $42.02802$\\ 
 $\Omega_c$ [deg] & $17.33160$ & $2.58516$ & $8.69490$\\ 
 $M_c$ [deg] & $109.04706$ & $199.56426$ & $142.82642$\\ 
 \hline
 $\rho_b$ [g cm$^{-3}$] & $2.3$ & $2.8$ & $2.4$\\ 
 $\rho_c$ [g cm$^{-3}$] & $0.9$ & $2.7$ & $2.9$\\ 
 $\rho_d$ [g cm$^{-3}$] & $2.2$ & $3.9$ & $4.2$\\ 
 $i_\mathrm{mut,bc}$ [deg] & $6.3$ & $17.6$ & $25.0$\\ 
 $i_\mathrm{mut,cd}$ [deg] & $7.6$ & $4.6$ & $8.0$\\ 
 $i_\mathrm{mut,de}$ [deg] & $20.2$ & $27.7$ & $48.2$\\ 
\hline
$\chi^2$ & $477.3$ & $502.0$ & $502.6$\\ 
$N_\mathrm{dof}$ & $435$ & $435$ & $435$\\ 
\hline
\end{tabular}
\label{table:Kepler411_solns}
\tablecomments{The $1$:$2$ solution was reported by \citet{Sun2019}, and we found other plausible solutions.}
\end{table}

\begin{figure*}
\centering
\includegraphics[width=0.975\textwidth]{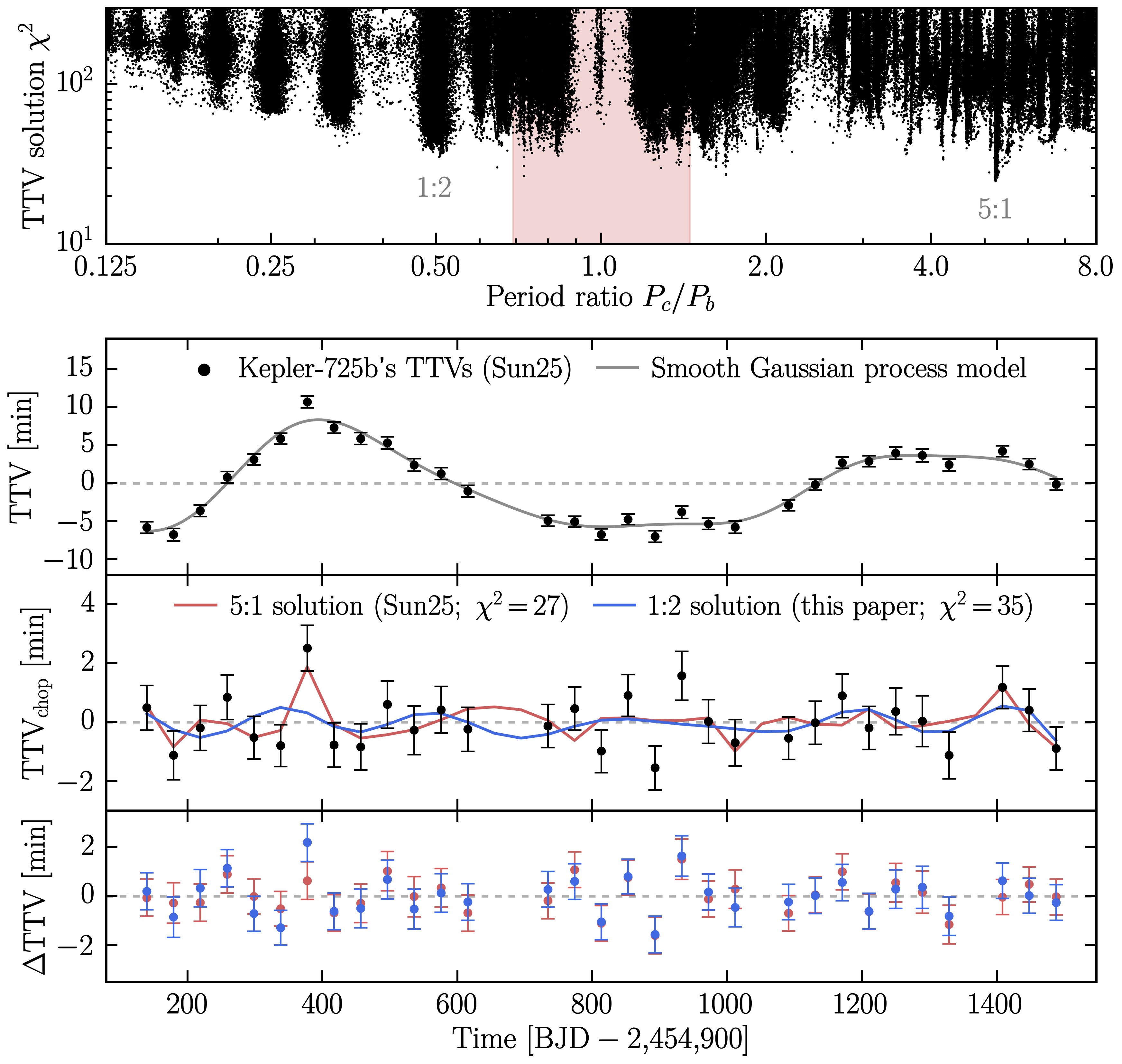}
\caption{{\bf Kepler-725}. {\it Top panel:} $\chi^2$ versus period ratio ($N_\mathrm{dof}\,{=}\,18$). The $5$:$1$ solution provides the best fit by a slim margin, followed by the $1$:$2$ solution and many others. {\it Second panel:} TTVs, as measured by \citet{Holczer2016} and used in \citet{Sun2025}. The gray curve is a GP model, used as a smooth interpolant to separate long- and short-timescale variations. {\it Third panel:} High-pass-filtered TTVs obtained by subtracting the GP. The red curve shows the best variant of the $5$:$1$ solution \citep{Sun2025}, and the blue curve shows the $1$:$2$ solution. {\it Fourth panel:} Residuals of the $5$:$1$ solution ($\mathrm{RMS}\,{=}\,0.71$\,min) and the $1$:$2$ solution ($\mathrm{RMS}\,{=}\,0.81$\,min). The differences between the models are minor; Kepler-725's TTV solution is not unique, and does not necessarily involve a planet in the habitable zone.}
\label{fig:Kepler725_TTVs}
\end{figure*}

After excluding solutions with physically unrealistic densities (greater than that of lead) and mutual inclinations between transiting planets exceeding $30^\circ$, the lowest-$\chi^2$ solution features a perturber with a 31-day period, near the $1$:$2$ MMR with planet d (Table~\ref{table:Kepler411_solns}). This solution resembles that reported by \citet{Sun2019}, although our inferred planet masses are somewhat lower. The model reproduces planet c's TTVs well, and the fit to planet d's TTVs remains satisfactory (Figure~\ref{fig:Kepler411_TTVs}; the data for planet b are not shown because there are no significant TTVs). It is worth noting that \citet{Sun2019} inflated the timing uncertainties of planet d by approximately a factor of $2$, whereas our adopted inflation corresponds to a $5$\% increase.

Alternative solutions can be found that are both physically plausible and fit the TTV data well. For instance, a $6$:$1$ configuration with a $2600~M_\oplus$ perturbing planet on a $370$-day orbit provides a comparably good fit ($\chi^2\,{=}\,502$; see Table~\ref{table:Kepler411_solns}). This model improves the fit to planet d's TTVs at the expense of a somewhat poorer fit to planet c. A $4$:$5$ configuration also produces a statistically similar fit, although it requires a moderately inclined perturber. The nominal $\chi^2$ difference of $25$ between the $1$:$2$ solution and the next-best alternatives might seem significant, but given the uncertainty in the timing uncertainties and the star's intrinsic variability, this distinction is not robust.

We conclude that the Kepler-411 system admits multiple physically plausible TTV solutions of comparable quality. The available TTV data, especially after allowing for modest underestimates of timing uncertainties, do not pinpoint a unique solution.

\subsection{Kepler-725: The Amphibian of TTVs}
\label{sec:Kepler725}

Kepler-725 (KOI-918) is a G-type star with $V\,{=}\,15.3$, mass
$0.95~M_\odot$, and radius $0.88~R_\odot$ \citep{Sun2025}. A single transiting planet is known, with a $40$-day period and a radius of $0.9~R_\mathrm{J}$ \citep{Borucki2011, Morton2016}. The planet has long been known to exhibit TTVs \citep{Ford2012a, Holczer2016}. \citet{Sun2025} modeled these variations and reported a unique solution: a companion with a mass of $10~M_\oplus$ on a $208$-day orbit near the $5$:$1$ MMR with planet b. Because such a planet would be located in the star's habitable zone, the uniqueness claim is of heightened interest.

We adopted the transit times and durations from the catalog of \citet{Holczer2016}, which were also used by \citet{Sun2025}, and performed the TTV solution search as described in Section~\ref{sec:TTV_modeling}. To enable a direct comparison with \citet{Sun2025}, we initially used only the TTV data, although we have confirmed that including TDVs leads to similar conclusions. The TTV solution search grid and two representative solutions are depicted in Figure~\ref{fig:Kepler725_TTVs}.

Our search recovered solutions consistent with the reported $5$:$1$ configuration. One class of solutions invokes a ${\approx}\,15\,M_\oplus$ perturber on a moderately eccentric orbit ($e\,{\approx}\,0.4$) that is modestly inclined ($i_\mathrm{mut}\,{\approx}\,20^\circ$), in good agreement with \citet{Sun2025}. However, other variants fit the data equally well  ($\Delta \chi^2\,{\approx}\,2$), including configurations with higher masses and larger mutual inclinations. Model parameters are provided in Table~\ref{table:Kepler725_solns}. The nominal best-fitting $5$:$1$ solution yields $\chi^2\,{=}\,27$ for $18$ degrees of freedom ($31\,{-}\,13$), suggesting (as usual) that the timing uncertainties may be somewhat underestimated.

We also identified alternative families of solutions that provide comparably good fits. In particular, an $8\,M_\oplus$ perturber with a $20$-day orbit, near the $1$:$2$ MMR, reproduces the observed TTVs with $\chi^2\,{=}\,35$. The perturber's low eccentricity ($e_c\,{\approx}\,0.1$) and lower-order MMR make this solution seem more typical of other {\it Kepler} systems than the $5$:$1$ solution.

The difference in $\chi^2$ between the $5$:$1$ and $1$:$2$ solutions is $8$, which could formally be interpreted as mild evidence in favor of the $5$:$1$ solution. Such a conclusion depends sensitively on the assumed timing uncertainties. The transit times reported by \citet{Holczer2016} were derived using an automated procedure in which uncertainties were estimated from the inverse Hessian matrix of the transit model fit. While computationally efficient, this approach does not capture the full posterior structure and may underestimate errors, particularly for low signal-to-noise ratio transits or in the presence of strong parameter covariances. Additionally, the transit light curves of Kepler-725\,b exhibit spot-crossing events, which can bias the transit times if not modeled explicitly.

\begin{figure*}
\centering
\includegraphics[width=0.975\textwidth]{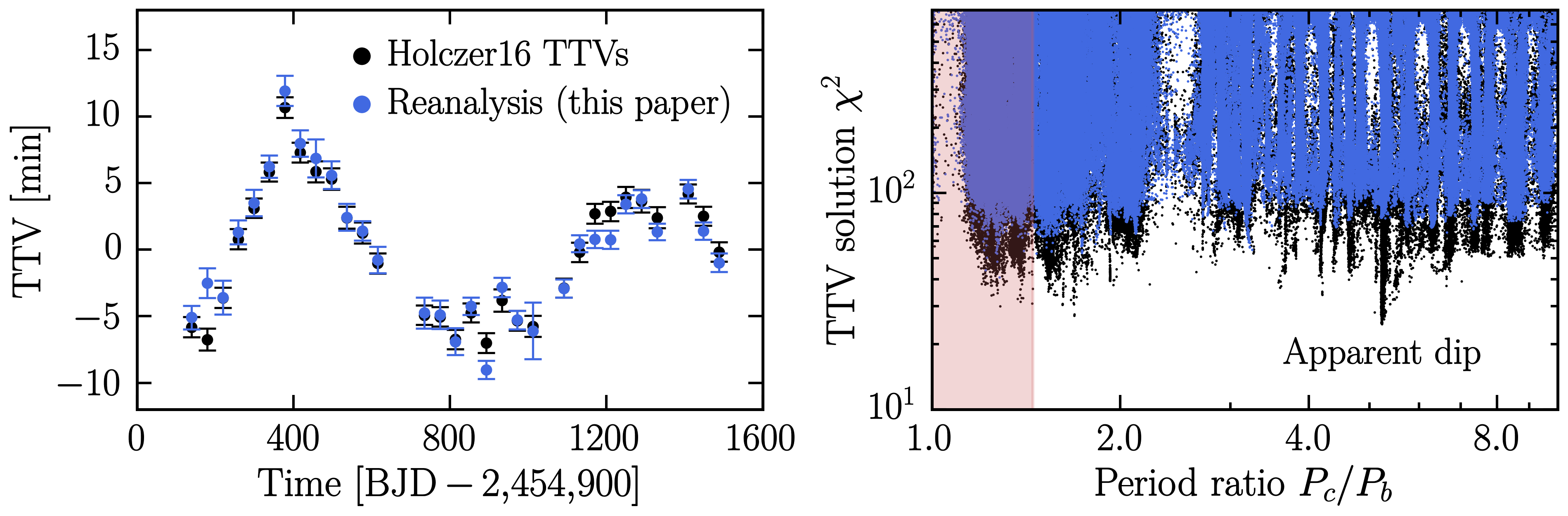}
\caption{{\bf Kepler-725}. {\it Left panel:} TTVs from the catlog of \citet{Holczer2016} and from our reanalysis of the {\it Kepler} data. {\it Right panel:} $\chi^2$ versus period ratio for searches performed on the two sets of TTV data. Our more deliberate analysis of the transits led to slight modifications in the times and uncertainties. Although small, these changes significantly affect the TTV solution landscape, invalidating the mild preference for the $5$:$1$ solution \citep{Sun2025}. More broadly, caution is advised when interpreting mild preferences, due to the sensitivity of TTV solutions to timing outliers and uncertainties.}
\label{fig:Kepler725_my_TTVs}
\end{figure*}

\begin{table}
\centering
\caption{{\bf Kepler-725}. Parameters for Two Models Fit to the TTV Data}
\begin{tabular}{ccc}
 \hline
 Parameter & $5$:$1$ Solution & $1$:$2$ Solution\\
 \hline
 $m_b$ [$M_\oplus$] & $521.76244$ & $666.95412$\\ 
 $P_b$ [days] & $39.6433684$ & $39.6486749$\\ 
 $e_b \cos(\omega_b)$ & $0.0131387$ & $-0.0706007$\\ 
 $e_b \sin(\omega_b)$ & $-0.1209431$ & $0.0479895$\\ 
 $i_b$ [deg] & $89.18745$ & $89.50101$\\ 
 $M_b$ [deg] & $344.50549$ & $124.32917$\\ 
\hline
 $m_c$ [$M_\oplus$] & $16.62026$ & $7.80208$\\ 
 $P_c$ [days] & $207.8208618$ & $20.0753096$\\ 
 $e_c \cos(\omega_c)$ & $0.3602292$ & $-0.1207585$\\ 
 $e_c \sin(\omega_c)$ & $0.1981198$ & $0.0535076$\\ 
 $i_c$ [deg] & $89.69658$ & $84.51226$\\ 
 $\Omega_c$ [deg] & $22.51236$ & $-16.54761$\\ 
 $M_c$ [deg] & $73.98472$ & $40.59120$\\ 
\hline
 $\rho_b$ [g cm$^{-3}$] & $2.3$ & $2.9$\\ 
 $i_\mathrm{mut}$ [deg] & $22.5$ & $17.3$\\ 
\hline
$\chi^2$ & $26.9$ & $34.9$\\ 
$N_\mathrm{dof}$ & $18$ & $18$\\ 
\hline
\end{tabular}
\label{table:Kepler725_solns}
\tablecomments{The $5$:$1$ solution was reported by \citet{Sun2025}, and we found the alternative $1$:$2$ solution.}
\end{table}

To assess the robustness of the published timing data, we independently re-derived the transit times using the process described in Section~\ref{sec:KOI142}. Prior to fitting, we masked three prominent starspot-crossing features. We stacked the resulting normalized light curves and measured the following parameters: $a/R_\star\,{=}\,49.40$, $R_p/R_\star\,{=}\,0.1135$, $b\,{=}\,0.3050$, $u_1\,{=}\,0.5073$, and $u_2\,{=}\,0.1786$. Our re-derived TTVs are shown in the left panel of Figure~\ref{fig:Kepler725_my_TTVs}. They differ somewhat from those of \citet{Holczer2016}, in $12$ cases by more than $1\sigma$ relative to the previously quoted uncertainties. These differences materially affect the TTV modeling. When we repeated the TTV solution search with our revised transit times, the $5$:$1$ solution was no longer preferred (see the right panel of Figure~\ref{fig:Kepler725_my_TTVs}). Instead, a $3$:$1$ model emerged as the best plausible solution, closely followed by $2$:$1$ and $6$:$1$ solutions. This reinforces the conclusion that Kepler-725's TTV solution is not unique, and serves as a cautionary illustration that mild preferences often depend sensitively on timing outliers and uncertainties.

Kepler-725 has multiple viable TTV solutions, and the preference among them depends on the details of the timing extraction and uncertainty model, for which confidence is limited by the star's spot activity. Thus, it is unclear whether Kepler-725\,c is hot and dry or wet and potentially habitable.

\begin{figure*}
\centering
\includegraphics[width=0.95\textwidth]{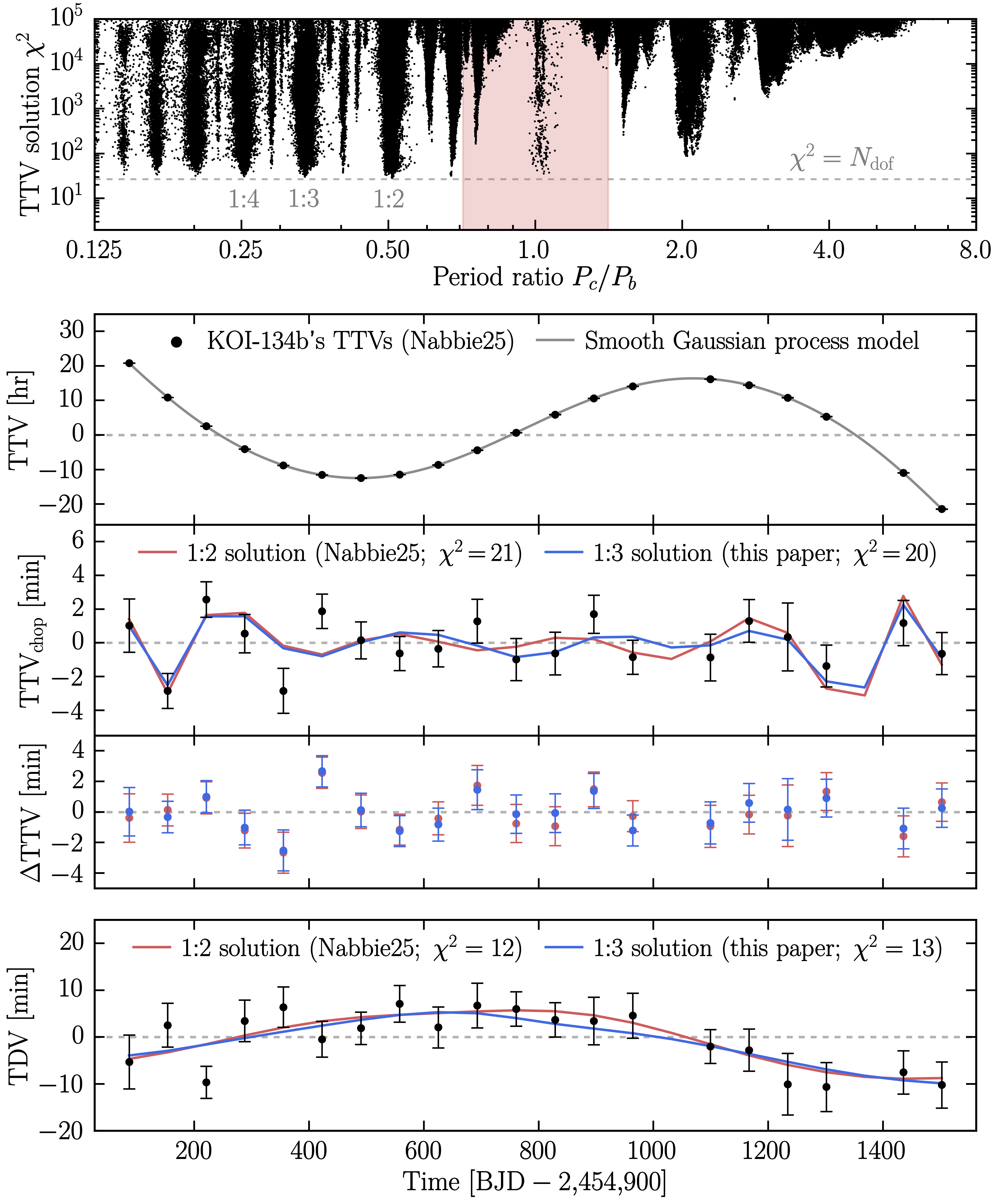}
\caption{{\bf KOI-134}. {\it Top panel:} $\chi^2$ versus period ratio ($N_\mathrm{dof}\,{=}\,27$). Many $1$:$N$ solutions provide a satisfactory fit to the data. {\it Second panel:} TTVs from \citet{Nabbie2025}. The gray curve is a GP model, used as a smooth interpolant to separate long- and short-timescale variations. {\it Third panel:} High-pass-filtered TTVs obtained by subtracting the GP. The red curve shows the best variant of the $1$:$2$ solution \citep{Nabbie2025}, and the blue curve shows the $1$:$3$ solution. {\it Fourth panel:} Residuals of the $1$:$2$ solution ($\mathrm{RMS}\,{=}\,1.2$\,min) and the $1$:$3$ solution ($\mathrm{RMS}\,{=}\,1.1$\,min). {\it Fifth panel:} TDVs and predictions of the $1$:$2$ and $1$:$3$ models. Both models fit the data well; KOI-134's TTV solution is not unique, and does not necessarily involve a companion with a large mutual inclination.}
\label{fig:KOI134_TTVs}
\end{figure*}

\subsection{KOI-134: The Sphinx of TTVs}
\label{sec:KOI134}

KOI-134 is an F-type star with $V\,{=}\,14$, mass $1.407~M_\odot$, and radius $1.667~R_\odot$ \citep{Nabbie2025}. The transiting-planet candidate KOI-134\,b was initially flagged by the {\it Kepler} pipeline but classified as a false positive because its large TTVs ($20$~hours; see Figure~\ref{fig:KOI134_TTVs}) caused it to fail the ``model-shift uniqueness'' test. This check searches for dips of similar depth and duration to the detected transit signal that could indicate problems with the light curve. Because the TTV amplitude greatly exceeds the transit duration, genuine transits were misidentified as spurious dips \citep{Coughlin2016, Nabbie2025}. The object was subsequently reinstated as a planet candidate \citep{Bryson2017} and later confirmed \citep{Nabbie2025}. The system's TTVs and TDVs were modeled by \citet{Nabbie2025}, who reported a unique solution in which a ${\approx}\,70\,M_\oplus$ companion resides on a $34$-day orbit near the $1$:$2$ MMR, with a mutual inclination of ${\approx}\,15^\circ$.

We adopted the transit times and durations from \citet{Nabbie2025} and carried out our TTV solution search (Section~\ref{sec:TTV_modeling}). Our search identified many viable solutions, mostly with the perturber interior to the transiting planet. The $\chi^2$ landscape  (Figure~\ref{fig:KOI134_TTVs}, top panel) shows multiple solution families clustered near $1$:$N$ commensurabilities.

Many of the lowest-$\chi^2$ solutions require high eccentricities ($e\,{\approx}\,0.5$), such that the perturber's apastron distance exceeds the transiting planet's periastron distance. These configurations are likely unstable and were excluded from further consideration. Among the remaining solutions, some involve severely misaligned orbits ($i_\mathrm{mut}\,{\gtrsim}\,60^\circ$), but there are also lower-inclination solutions orbits ($i_\mathrm{mut}\,{\lesssim}\,20^\circ$) that provide statistically similar fits ($\Delta \chi^2\,{\lesssim}\,5$). In the best-fitting nearly coplanar configuration, the perturber is near the $1$:$3$ MMR with a mass of ${\approx}\,90~M_\oplus$ (Table~\ref{table:KOI134_solns}). The inferred bulk density ($1.6$\,g\,cm$^{-3}$), moderate eccentricities ($e\,{\approx}\,0.3$), and modest mutual inclination ($15^\circ$) are all plausible.

The search also turned up $1$:$2$ solutions that fit the data equally well. One such solution involves a more massive perturber (${\approx}\,2.5~M_\mathrm{J}$) and nearly circular orbits for both planets ($e\,{<}\,0.05$), with a mutual inclination of $15^\circ$.  \citet{Nabbie2025} also reported a $1$:$2$ solution, although with somewhat different parameters: moderate eccentricities ($e_b\,{\approx}\,0.16$ and $e_c\,{\approx}\,0.24$), a lower perturber mass ($m_c\,{\approx}\,70\,M_\oplus$), and a similar mutual inclination. Their reported reduced $\chi^2$ value of $1.52$ corresponds to $\chi^2\,{=}\,41$, assuming $N_\mathrm{dof}\,{=}\,27$, while our $1$:$2$ solution provides a marginally improved fit ($\Delta \chi^2\,{=}\,7$) with lower eccentricities.

Importantly, the transit-timing data do not demand a substantially inclined perturber. We found multiple $1$:$2$, $1$:$3$, and $1$:$4$ solutions with mutual inclinations below $10^\circ$. In one $1$:$4$ configuration (listed in Table~\ref{table:KOI134_solns}), the mutual inclination is only $2.6^\circ$, implying that the companion is likely to transit. Other low-inclination solutions exist in which the geometry would not lead to observable transits.

Despite having the highest signal-to-noise ratio in our sample, the TTVs of KOI-134\,b have no unique solution. They admit a broad range of dynamically plausible solutions spanning multiple commensurabilities and inclination regimes. The data do not single out a unique configuration, nor do they strongly constrain the inclination or mass of the companion.

\subsection{Kepler-138: The Shape-shifter of TTVs}
\label{sec:Kepler138}

Kepler-138 is a relatively bright ($V\,{=}\,13$) M-dwarf with mass $0.535~M_\odot$ and radius $0.535~R_\odot$ \citep{Piaulet2023}. It harbors a compact system of three transiting planets, with periods of $10.3$, $13.8$, and $23.0$~days and radii of $0.6$, $1.5$, and $1.5~R_\oplus$ \citep{Borucki2011, Rowe2014}. \citet{Kipping2014} modeled the TTVs of planets c and d to determine their masses. With the full {\it Kepler} dataset, TTVs of planet b were detected, and all three masses were constrained \citep{JontofHutter2015}. More recently, \citet{Piaulet2023} reported $13$ additional transit times of planet d based on observations with the {\it Hubble} and {\it Spitzer} space telescopes. The new timings deviated significantly from the predictions of three-planet dynamical models, and no satisfactory three-planet solution could be found. They concluded that a fourth planet of mass ${\approx}\,0.4~M_\oplus$ on a $38$-day orbit was required.

We collected the {\it Kepler} transit times from \citet{JontofHutter2015} and the {\it HST}/{\it Spitzer} transits from \citet{Piaulet2023}. We first attempted to model the system using only the three transiting planets. The best-fit models required extreme mutual inclinations (${\gtrsim}\,80^\circ$), which is geometrically unlikely. Nearly coplanar solutions demanded unrealistically large masses for planet c (${\gtrsim}\,15\,M_\oplus$), corresponding to bulk densities far exceeding that of iron. The best plausible three-planet solution (with $i_\mathrm{mut}\,{\lesssim}\,15^\circ$ and reasonable densities) could do no better than $\chi^2\,{=}\,794$ for $250$ degrees of freedom ($270\,{-}\,20$). In light of these results, we concur that a three-planet model cannot adequately explain the transit-timing data.

We continued the search after adding a fourth planet to the model. The plot of $\chi^2$ versus period (Figure~\ref{fig:Kepler138_TTVs}, top panel) shows a wide range of possible solutions, with perturber periods ranging from approximately $5$ to $200$~days and masses from sub-Earth up to hundreds of Earth masses. The improvement relative to the best three-planet model is substantial ($\Delta \chi^2\,{\approx}\,300$), justifying the seven additional free parameters ($\Delta \mathrm{BIC}\,{\approx}\,270$). However, even the best four-planet solutions provide a statistically unsatisfactory fit, with $\chi^2\,{\approx}\,480$ for $243$ degrees of freedom, likely reflecting underestimated timing uncertainties for the shallow transits of Kepler-138.

\begin{table}
\centering
\caption{{\bf KOI-134}. Parameters for Three Models Fit to the TTV and TDV Data}
\begin{tabular}{cccc}
 \hline
 Parameter & $1$:$2$ Solution & $1$:$3$ Solution & $1$:$4$ Solution\\
 \hline
 $m_b$ [$M_\oplus$] & $483.56509$ & $518.66069$ & $422.62600$\\ 
 $P_b$ [days] & $67.6167193$ & $67.1310773$ & $67.0543485$\\ 
 $e_b \cos(\omega_b)$ & $0.0183022$ & $0.1256345$ & $-0.2310348$\\ 
 $e_b \sin(\omega_b)$ & $-0.0402990$ & $-0.3240110$ & $-0.1535570$\\ 
 $i_b$ [deg] & $89.99522$ & $89.38284$ & $89.58008$\\ 
 $M_b$ [deg] & $51.20022$ & $36.99656$ & $162.95686$\\ 
 \hline
 $m_c$ [$M_\oplus$] & $804.06858$ & $89.60571$ & $87.00325$\\ 
 $P_c$ [days] & $33.2927047$ & $22.6393139$ & $16.8265362$\\ 
 $e_c \cos(\omega_c)$ & $-0.0284570$ & $0.1260321$ & $-0.5881505$\\ 
 $e_c \sin(\omega_c)$ & $-0.0399899$ & $-0.2700026$ & $0.2629920$\\ 
 $i_c$ [deg] & $69.26498$ & $104.28250$ & $90.41860$\\ 
 $\Omega_c$ [deg] & $-5.81842$ & $1.02547$ & $2.39619$\\ 
 $M_c$ [deg] & $188.17646$ & $74.12452$ & $312.54584$\\ 
 \hline
 $\rho_b$ [g cm$^{-3}$] & $1.5$ & $1.6$ & $1.3$\\ 
 $i_\mathrm{mut}$ [deg] & $21.5$ & $14.9$ & $2.5$\\ 
 \hline
 $\chi^2$ & $33.9$ & $33.4$ & $43.9$\\ 
 $N_\mathrm{dof}$ & $27$ & $27$ & $27$\\ 
 \hline
\end{tabular}
\label{table:KOI134_solns}
\tablecomments{The $1$:$2$ solution was reported by \citet{Nabbie2025}, and we found several other possible solutions.}
\end{table}

Many of the four-planet solutions are physically plausible, featuring low eccentricities, modest mutual inclinations, and reasonable bulk densities for the transiting planets. Table~\ref{table:Kepler138_solns} gives the parameters of three solutions: the $5$:$3$ solution reported by \citet{Piaulet2023}, a $1$:$3$ solution with an internal perturber, and a $4$:$1$ solution involving a more massive outer companion. As displayed in the lower panels of Figure~\ref{fig:Kepler138_TTVs}, the $5$:$3$ and $1$:$3$ solutions provide statistically similar fits, with a slight preference for the $1$:$3$ solution ($\Delta \chi^2\,{=}\,7$). Other MMRs produce comparably good fits.

\begin{figure*}
\centering
\includegraphics[width=0.95\textwidth]{Kepler138_TTVs.png}
\caption{{\bf Kepler-138}. {\it Top panel:} $\chi^2$ versus period ratio ($N_\mathrm{dof}\,{=}\,243$). Many solutions provide an equally good fit to the TTV data. {\it Second panel:} Planet b's TTVs, as measured by \citet{JontofHutter2015}. The red curve shows the best variant of the $5$:$3$ solution \citep{Piaulet2023}, and the blue curve shows the $1$:$3$ solution. {\it Third panel:} Planet c's TTVs, along with the predictions of the $5$:$3$ and $1$:$3$ models. {\it Fourth panel:} Planet d's TTVs, including {\it Kepler} transits from \citet{JontofHutter2015} and {\it HST}/{\it Spitzer} transits from \citet{Piaulet2023}. {\it Fifth panel:} Planet d's TTV residuals for the $5$:$3$ solution ($\mathrm{RMS}\,{=}\,5.0$\,min) and $1$:$3$ solution ($\mathrm{RMS}\,{=}\,5.0$\,min). These two models, and many others, fit Kepler-138's TTV data indistinguishably well ($\Delta \chi^2\,{=}\,7$).}
\label{fig:Kepler138_TTVs}
\end{figure*}

Thus, while the existence of a fourth planet is supported by the data, the specific architecture reported by \citet{Piaulet2023} is only one of many possibilities. The inferred masses of planets b, c, and d vary substantially across different four-planet solutions. Because the mass estimates are correlated with the assumed properties of the nontransiting companion, the previously published mass values and uncertainties are unlikely to be reliable.

\subsection{TOI-4562: The Rorschach Test of TTVs}
\label{sec:TOI4562}

TOI-4562 is the only TESS-discovered object in our sample. It is a fairly bright ($V\,{=}\,12$), young (${\sim}\,700$\,Myr) F-type star with mass $1.192\,M_\odot$ and radius $1.152\,R_\odot$ \citep{Heitzmann2023}. Using $25$ sectors of TESS photometry, \citet{Heitzmann2023} detected a transiting planet with a $225$-day orbital period and a large eccentricity ($e_b\,{=}\,0.76$). They reported TTVs with an amplitude of about $10$\,min. \citet{Fermiano2024} conducted ground-based observations of two additional transits and performed dynamical modeling using an MCMC code. They reported a unique solution for the companion: $P_c\,{\approx}\,3990$\,days and $m_c\,{\approx}\,5.77\,M_\mathrm{J}$.

\begin{figure}
\centering
\includegraphics[width=0.475\textwidth]{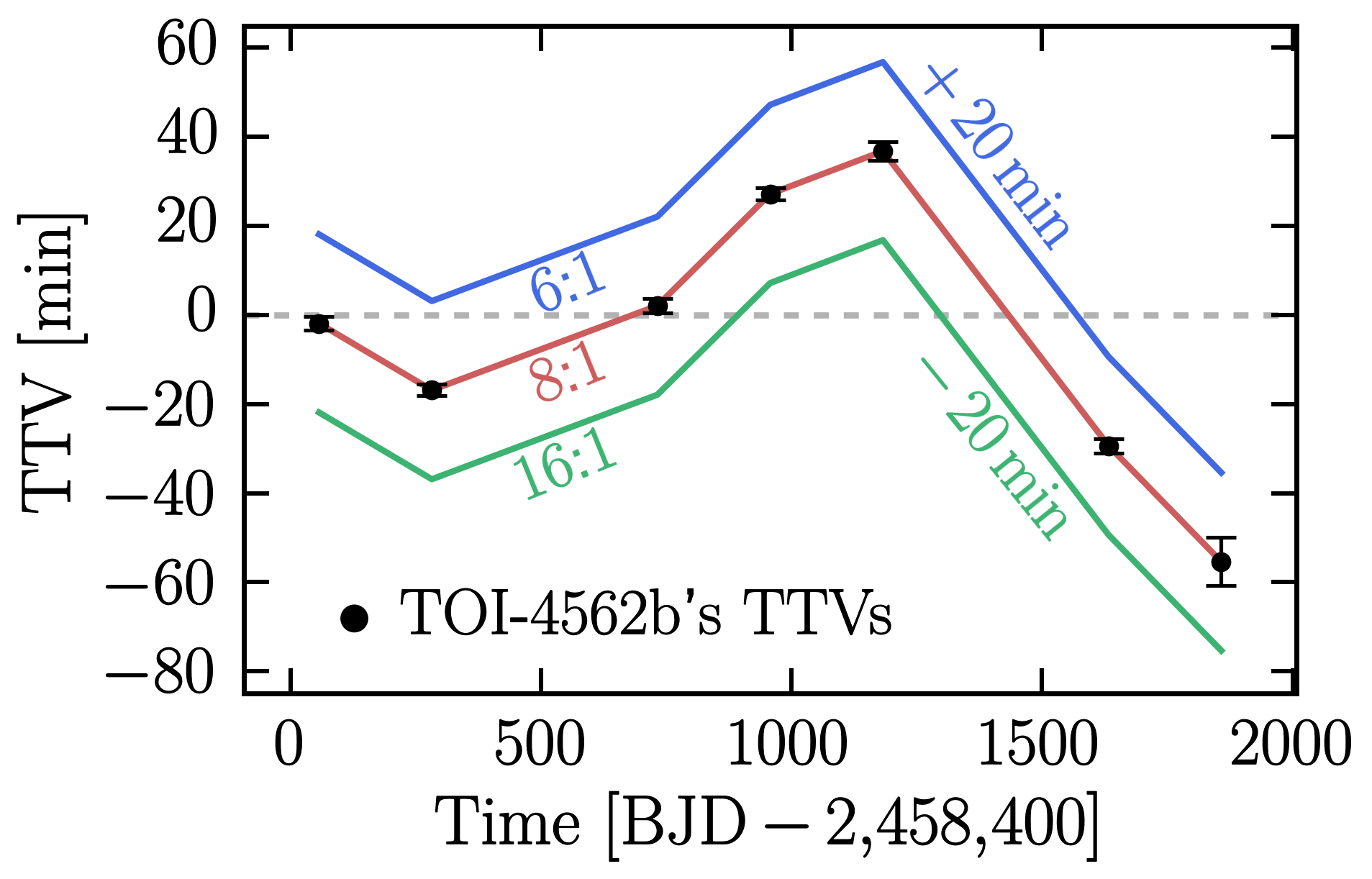}
\caption{TTVs of TOI-4562\,b, from \citet{Fermiano2024}. Colored lines show predictions of dynamical models with companions near different $N$:$1$ MMRs. For the purposes of visualization, the $6$:$1$ and $16$:$1$ models have been spaced vertically by arbitrary offsets of $20$~min. All models match the transit times to high precision ($\chi^2\,{<}\,10^{-3}$) despite having companions with very different periods ($1280$\,--\,$3521$\,days), masses ($146$\,--\,$1180\,M_\oplus$), and eccentricities ($0.15$\,--\,$0.46$). In this case, there is no unique solution to the TTV inversion problem.}
\label{fig:TOI4562}
\end{figure}

We adopted the published transit times and carried out our standard search process. With only seven transit-timing measurements and low-amplitude TTVs, the solution is poorly constrained (Figure~\ref{fig:TOI4562}). We found viable solutions spanning a wide range of planet masses and orbital periods ($50$ to $2000$~days). Many of these solutions reproduce the observed transit times essentially perfectly, with $\chi^2\,{\lesssim}\,10^{-3}$. These models are interpolating the data points rather than being constrained by them.

\begin{table}
\centering
\caption{{\bf Kepler-138}. Parameters for Three Models Fit to the TTV Data}
\begin{tabular}{cccc}
 \hline
 Parameter & $5$:$3$ Solution & $1$:$3$ Solution & $4$:$1$ Solution\\
 \hline
 $m_b$ [$M_\oplus$] & $0.18775$ & $0.38845$ & $0.10231$\\ 
 $P_b$ [days] & $10.3126859$ & $10.3160513$ & $10.3131546$\\ 
 $e_b \cos(\omega_b)$ & $0.0110584$ & $0.0625888$ & $0.0189651$\\ 
 $e_b \sin(\omega_b)$ & $-0.0805155$ & $0.0543634$ & $0.0039143$\\ 
 $i_b$ [deg] & $89.71845$ & $89.71666$ & $89.63495$\\ 
 $M_b$ [deg] & $7.81662$ & $239.76556$ & $273.06849$\\ 
\hline
 $m_c$ [$M_\oplus$] & $3.74726$ & $3.82385$ & $0.54911$\\ 
 $P_c$ [days] & $13.7807256$ & $13.7833528$ & $13.7806874$\\ 
 $e_c \cos(\omega_c)$ & $0.0156276$ & $0.0604308$ & $0.0150724$\\ 
 $e_c \sin(\omega_c)$ & $-0.0725284$ & $0.0435845$ & $-0.0052800$\\ 
 $i_c$ [deg] & $89.16341$ & $90.22446$ & $91.24043$\\ 
 $\Omega_c$ [deg] & $1.45218$ & $0.00004$ & $2.24007$\\ 
 $M_c$ [deg] & $150.15108$ & $31.70280$ & $91.74454$\\ 
\hline
 $m_d$ [$M_\oplus$] & $5.26805$ & $5.67577$ & $6.47660$\\ 
 $P_d$ [days] & $23.0921587$ & $23.0919774$ & $23.0922212$\\ 
 $e_d \cos(\omega_d)$ & $0.0255792$ & $0.0791945$ & $0.0230879$\\ 
 $e_d \sin(\omega_d)$ & $-0.0412489$ & $0.0419631$ & $0.0110025$\\ 
 $i_d$ [deg] & $89.64805$ & $89.74838$ & $88.89716$\\ 
 $\Omega_d$ [deg] & $0.00051$ & $7.23935$ & $3.70651$\\ 
 $M_d$ [deg] & $323.77233$ & $231.91508$ & $240.49905$\\ 
\hline
 $m_e$ [$M_\oplus$] & $1.60678$ & $3.50120$ & $150.35181$\\ 
 $P_e$ [days] & $38.2409969$ & $7.6484786$ & $94.1053801$\\ 
 $e_e \cos(\omega_e)$ & $0.0631076$ & $0.0874053$ & $0.0147595$\\ 
 $e_e \sin(\omega_e)$ & $-0.1039855$ & $0.0520244$ & $0.1815031$\\ 
 $i_e$ [deg] & $98.46699$ & $80.11268$ & $93.41138$\\ 
 $\Omega_e$ [deg] & $14.57688$ & $-2.43533$ & $14.88069$\\ 
 $M_e$ [deg] & $289.59977$ & $332.43608$ & $160.61404$\\ 
\hline
 $\rho_b$ [g cm$^{-3}$] & $3.9$ & $8.1$ & $2.1$\\ 
 $\rho_c$ [g cm$^{-3}$] & $6.0$ & $6.1$ & $0.9$\\ 
 $\rho_d$ [g cm$^{-3}$] & $8.4$ & $9.1$ & $10.3$\\ 
 $i_\mathrm{mut,de}$ [deg] & $17.0$ & $9.9$ & $15.3$\\ 
\hline
$\chi^2$ & $491.4$ & $484.1$ & $483.3$\\ 
$N_\mathrm{dof}$ & $243$ & $243$ & $243$\\ 
\hline
\end{tabular}
\label{table:Kepler138_solns}
\tablecomments{The $5$:$3$ solution was reported by \citet{Piaulet2023}, and we found many other possible solutions.}
\end{table}

\citet{Heitzmann2023} reported constraints on $m_b$ and $e_b$ that we did not enforce during the fitting process. Nonetheless, some solutions happen to satisfy these constraints. Figure~\ref{fig:TOI4562} shows three models that fit the TTV data ``perfectly,'' while also satisfying the $1\sigma$ constraints on  $m_b$ and $e_b$ from \citet{Heitzmann2023}. In all three cases, the perturber's mutual inclination is ${\lesssim}\,15^\circ$. Yet, the properties of the perturbing planets in the three solutions are dramatically different, with $P_c\,{=}\,1280$\,--\,$3521$\,days, $m_c\,{=}\,146$\,--\,$1180\,M_\oplus$, and $e_c\,{=}\,0.15$\,--\,$0.46$.

With so few data points, many different configurations can reproduce the timing measurements to high precision. Because of the nearly exact degeneracy between different solution families, the companion properties reported by \citet{Fermiano2024} are not well constrained by the available data.

\subsection{WASP-18: Insufficient evidence for TTVs}
\label{sec:WASP18}

WASP-18 is the brightest star in our sample, with $V\,{=}\,9.3$. It is an F-type star with mass $1.3~M_\odot$ and radius $1.3~R_\odot$ that hosts an ultra-hot Jupiter, WASP-18\,b, with a period of $0.94$~day, a mass of $10\,M_\mathrm{J}$, and a radius of $1.2\,R_\mathrm{J}$ \citep{Hellier2009, Southworth2009, CortesZuleta2020}. The system has been monitored extensively to search for evidence of orbital decay, apsidal precession, and additional companions \citep{Wilkins2017, McDonald2018, Csizmadia2019, Shporer2019, Maciejewski2020, Patra2020, Maciejewski2022}. Based on two sectors of TESS data, \citet{Pearson2019} reported short-timescale (${\approx}\,8$\,day), low-amplitude (${\approx}\,15$\,sec) quasiperiodic TTVs (see their Figure~15). Using a machine-learning model trained on $N$-body simulations, they inferred a unique solution involving a ${\approx}\,55~M_\oplus$ companion near the $2$:$1$ MMR with WASP-18\,b. Few hot Jupiters are known to have such close companions, making the proposed detection potentially important. However, the reported TTV amplitude is comparable to the timing uncertainties, warranting careful scrutiny.

We re-analyzed the available TESS photometry using the procedure described in Section~\ref{sec:KOI142}. Since the study by \citet{Pearson2019}, WASP-18 has been observed in four additional TESS sectors ($29$, $30$, $69$, and $96$). A total of $137$ transits were analyzed, after excluding partial transits. The parameters that best fit the stacked light curve are  $a/R_\star\,{=}\,3.449$, $R_p/R_\star\,{=}\,0.09847$, $b\,{=}\,0.3908$, $u_1\,{=}\,0.2920$, and $u_2\,{=}\,0.1646$.

The transit times for the first two TESS sectors, as analyzed by \citet{Pearson2019}, are shown in the top panel of Figure~\ref{fig:WASP18_126}. When restricting attention to this data, we found a constant-period model to be an adequate fit, with $\chi^2\,{=}\,42.9$ for $44$ degrees of freedom ($46\,{-}\,2$), and $\mathrm{BIC}\,{=}\,50.5$. Adding a sinusoidal function (three additional free parameters) reduces $\chi^2$ to $36.5$ for $41$ degrees of freedom, but increases BIC to $55.6$. The $\Delta$BIC of $5.1$ suggests that the modest reduction in $\chi^2$ does not justify the additional parameters. Furthermore, the model did not turn out to have predictive power. When predictions across all available TESS sectors were included, the preference for the constant-period model grew ($\Delta \mathrm{BIC}\,{=}\,10$). For this comparison, we estimated the uncertainties on the two models' predictions using the inverse of the Hessian matrix. Refitting the sinusoidal model to all the data led to a somewhat different best-fit period and phase, and it remained disfavored relative to the refitted constant-period model ($\Delta \mathrm{BIC}\,{=}\,1.0$). Dynamical models fit to the TTV data are disfavored even more strongly by the BIC because of the additional degrees of freedom.

We therefore concluded that the existence of WASP-18\,c is not supported by the current transit-timing data. The available data do not provide statistically robust evidence for TTVs of WASP-18\,b, nor for any companion.

\subsection{WASP-126: Insufficient evidence for TTVs}
\label{sec:WASP126}

\begin{figure}
\centering
\includegraphics[width=0.475\textwidth]{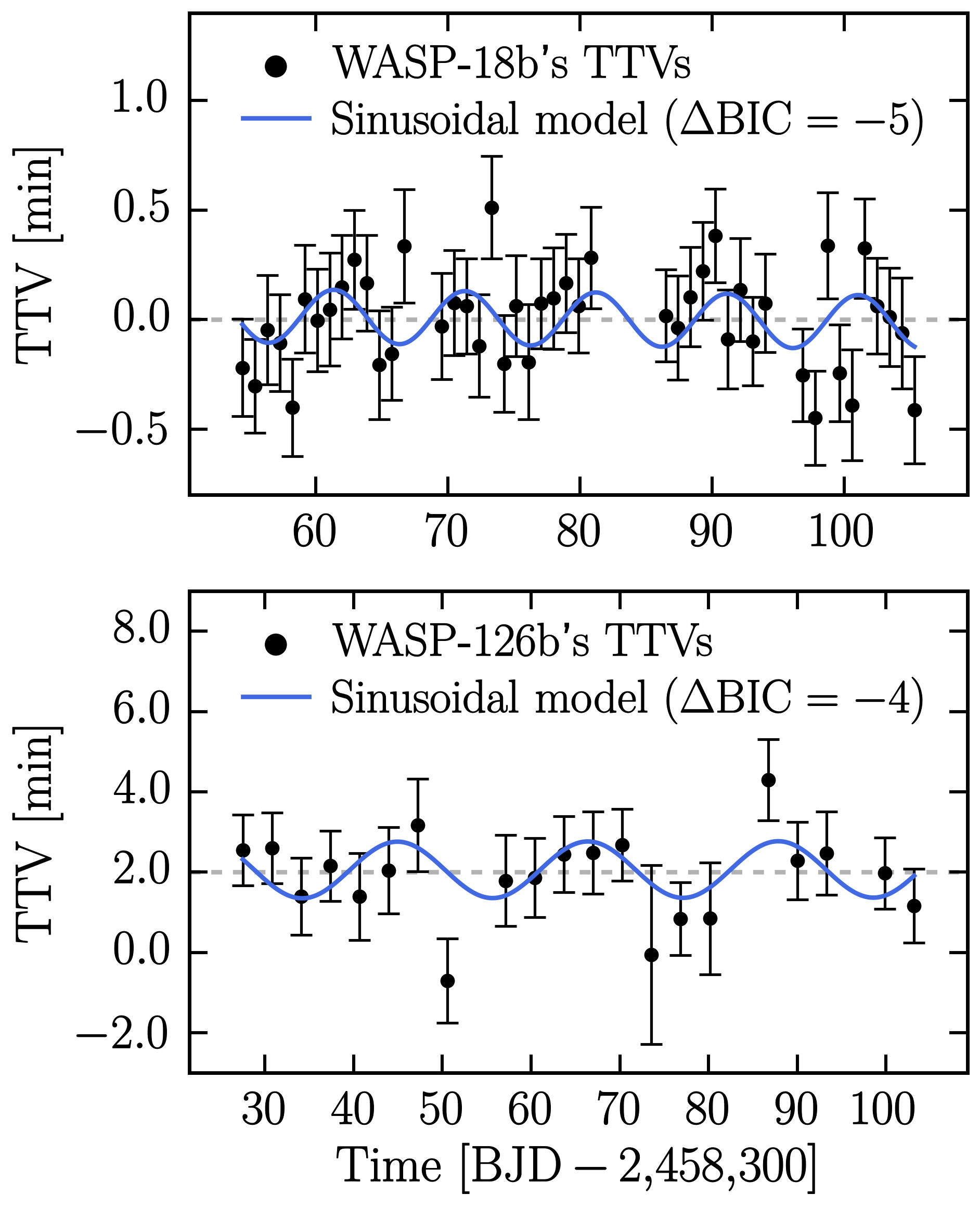}
\caption{TTVs from TESS sectors $1$\,--\,$3$ for WASP-18b (top) and WASP-126b (bottom). In both cases, the TTVs appear approximately flat, and a linear ephemeris is preferred to the sinusoidal TTV model according to the BIC statistic. The sinusoidal models also generalize poorly to the more recent TESS data, with $\Delta \mathrm{BIC}\,{\approx}\,10$ between the predicted transit times of the linear and sinusoidal models for both WASP-18\,b and WASP-126\,b. The TTV data do not support the existence of WASP-18\,c or WASP-126\,c \citep{Pearson2019}.}
\label{fig:WASP18_126}
\end{figure}

WASP-126 is a relatively bright ($V\,{=}\,11.0$) G-type star with mass $1.1~M_\odot$ and radius $1.3~R_\odot$. A hot Jupiter orbits the star with period $3.3$~days, mass $0.3~M_\mathrm{J}$, and radius $0.66~R_\mathrm{J}$ \citep{Maxted2016}. The implied bulk density of $0.4$\,g\,cm$^{-3}$ places WASP-126\,b in the category of inflated hot Jupiters. There have not been many published transit-timing studies of this system. Using three sectors of TESS data, \citet{Pearson2019} reported short-timescale (${\approx}\,7$\,day), low-amplitude (${\approx}\,1$\,min) quasiperiodic TTVs (see their Figure~12). They performed dynamical modeling and stated that the data uniquely identify the perturber as having a mass of ${\approx}\,60\,M_\oplus$ and a period that places it near the $2$:$1$ MMR. As with WASP-18\,b, the candidate TTV signal has a low signal-to-noise ratio.

WASP-126 lies near the southern ecliptic pole and has therefore been observed in numerous TESS sectors; the current total is $41$. We downloaded all the available data from MAST and performed a uniform transit analysis following the procedure described in Section~\ref{sec:KOI142}. After excluding partial transits, we were left with $273$ TESS transits. We derived the following best-fit parameters from the stacked light curve: $a/R_\star\,{=}\,3.449$, $R_p/R_\star\,{=}\,0.09847$, $b\,{=}\,0.3908$, $u_1\,{=}\,0.2920$, and $u_2\,{=}\,0.1646$.

To begin, we restricted the analysis to the same three sectors examined by \citet{Pearson2019}. The bottom panel of Figure~\ref{fig:WASP18_126} shows the residuals of the best-fit constant-period model ($\chi^2\,{=}\,19.7$, $N_\mathrm{dof}\,{=}\,19$, and $\mathrm{BIC}\,{=}\,25.8$) and those of the best-fit sinusoidal model ($\chi^2\,{=}\,14.7$, $N_\mathrm{dof}\,{=}\,16$, $\mathrm{BIC}\,{=}\,30.0$). The constant-period model is favored by  $\Delta \mathrm{BIC}\,{=}\,4.1$. As was the case with WASP-18, the sinusoidal model did not succeed at predicting the transit times observed in subsequent sectors, and was disfavored by $\Delta \mathrm{BIC}\,{=}\,8.5$ relative to the constant-period model. When both models were refitted to all the data, the sinusoidal model is disfavored by $\Delta \mathrm{BIC}\,{=}\,4.3$.

We therefore conclude that the existence of WASP-126\,c is not substantiated by the current data. Even with $273$ transits spanning $41$ sectors, the available transit-timing data provide no statistically robust evidence for TTVs of WASP-126\,b.

\section{Discussion}
\label{sec:discussion}

In this section, we list our conclusions for each of the $12$ TTV planets, discuss what separates systems with unique TTV solutions from the others, and summarize the lessons learned from our analysis.

\subsection{System summaries}
\label{sec:sys_summary}

The TTV inversion problem is subject to severe degeneracies. To assess how often unique solutions are truly obtained, we revisited the $12$ planets in the literature that were discovered and characterized using the transit-timing method. Our findings, ordered from most robust to least convincing, are summarized below.

\begin{itemize}

\item {\it KOI-142}: Although multiple near-resonant configurations reproduce the slow $600$-day TTV signal, only the $2$:$1$ solution first identified by \citet{Nesvorny2013} reproduces the conjunction-timescale structure. The system has a robust and unique TTV solution.

\item {\it Kepler-419}: A $700$-day, ${\approx}\,7\,M_\mathrm{Jup}$ companion is singled out as the only solution that provides a satisfactory fit to the data, as reported by \citet{Dawson2014}. Near-resonant alternatives are excluded due to the high precision of the {\it Kepler} transit times.

\item {\it KOI-872}: Several different configurations predict TTVs that provide good matches to the observed pattern, but the $5$:$3$ solution first identified by \citet{Nesvorny2012} is preferred by a large enough margin ($\Delta \chi^2\,{>}\,200$) to provide reasonable assurance that it is unique.

\item {\it KOI-884}: In addition to the $3$:$1$ solution advocated by \citet{Nesvorny2014}, we identified a $3$:$2$ solution that provides a better fit ($\Delta \chi^2\,{\approx}\,100$) and has physically plausible planet properties. This new solution narrowly exceeds the evidentiary threshold for uniqueness.

\item {\it Kepler-82}: Despite the detection of short-timescale TTVs by \citet{Freudenthal2019}, we found that the TTV solution is not unique. Two distinct configurations ($3$:$2$ and $3$:$1$) can explain the data equally well.

\item {\it Kepler-411}: We found at least two plausible families of TTV solutions that fit the data comparably well. The published solution, in which the perturber has a $31.5$-day period \citep{Sun2019}, is not uniquely preferred.

\item {\it Kepler-725}: The reported 208-day habitable-zone companion \citep{Sun2025} is only one of several possibilities. Alternative solutions fit the data nearly as well, including solutions with perturbers with periods as short as $25$~days.

\item {\it KOI-134}: The $1$:$2$ TTV solution described by \citet{Nabbie2025} is notable for its relatively high mutual inclination ($i_\mathrm{mut}\,{\approx}\,15^\circ$). However, several other $1$:$N$ solutions fit the TTV and TDV data at least as well, and in some of these solutions, the perturber's orbit is nearly coplanar with that of the transiting planet.

\item {\it Kepler-138}: We confirmed the claim by \citet{Piaulet2023} that a fourth planet is required to explain the TTV data, but we did not confirm that the system has a unique TTV solution. The properties of the fourth planet are very poorly constrained.

\item {\it TOI-4562}: With only seven transit times, numerous dynamically distinct models can match the data to high precision. The TTV solution offered by \citet{Fermiano2024} is highly nonunique.

\item {\it WASP-18}: Analysis of an expanded TESS dataset showed no compelling evidence of deviations from a constant-period model, nor for the perturbing planet proposed by \citet{Pearson2019}.

\item {\it WASP-126}: Similar to the case of WASP-18, our investigation of all the available TESS data did not provide evidence for timing anomalies, casting serious doubt on the existence of the planet WASP-126\,c \citep{Pearson2019}.

\end{itemize}

\subsection{Conditions required for a unique solution}
\label{sec:unique_conditions}

Detecting relatively fast structure in TTVs appears to be the most effective way to break degeneracies among competing families of solutions \citep[see, e.g.,][]{Nesvorny&Morbidelli2008, Veras2011, Boue2012, Deck&Agol2015}. These fast TTVs can be due either to the synodic chopping effect of two planets with low-eccentricity orbits --- as in KOI-142, KOI-872, and KOI-884 --- or strongly varying distances of closest approach due to large eccentricity, as in Kepler-419. In each of these cases, we found multiple solutions that can reproduce the super-period TTV signal; most of the discriminatory power that enables unique solutions comes from the faster-timescale TTVs.

The required signal-to-noise ratio can be substantial. For instance, if KOI-872's timing uncertainties had been larger by a factor of four, the two leading solutions shown in Figure~\ref{fig:KOI872_TTVs} would have differed by only $\Delta\chi^2\,{\approx}\,10$ and would not have been distinguishable. Conversely, systems lacking confidently detected fast TTVs admitted multiple solution families, even when the super-period TTV signal was detected at high significance. KOI-134 is an instructive example. Despite having TTVs detected with a higher signal-to-noise ratio than any of the systems with unique solutions, KOI-134 lacks a unique solution.

KOI-884 is an interesting borderline case that illustrates how sampling and aliasing can complicate the interpretation of conjunction-timescale TTV signals. In addition to the ${\approx}\,800$-day super-period signal, the TTVs of KOI-884\,b show a ${\approx}\,60$-day component (Figure~\ref{fig:KOI884_TTVs}). Yet, despite this clear detection of fast TTVs, both the $3$:$2$ and $3$:$1$ solutions fit the TTV data well. In the $3$:$2$ family, the $60$-day period corresponds directly to the synodic period, $(P_d^{-1} - P_e^{-1})^{-1}$. In the $3$:$1$ solution, the synodic period is $30$~days, which seems inconsistent with the fast TTV signal at first glance. However, TTVs are measured only at successive transits of the perturbed planet, which occur at intervals of $P_d\,{\approx}\,20.5$\,days. As dictated by the Nyquist limit, periodicities shorter than about $2 P_d\,{\approx}\,41$\,days cannot be uniquely resolved and may appear at aliased periods \citep{Yahalomi2025, Yahalomi+2026}. In this case, the dominant alias of the $30$-day synodic period falls near $60$~days, allowing the $3$:$1$ configuration to reproduce the observed fast TTV signal.

Similar degeneracies arose elsewhere in our sample. For KOI-872, a fast TTV signal with an $80$-day period is observed (Figure~\ref{fig:KOI872_TTVs}). In the family of $5$:$3$ solutions, this period corresponds to the synodic timescale; in the $5$:$2$ family, it is identified with the perturber's orbital period; and in the $3$:$5$ and $2$:$5$ solutions, it matches an alias of the synodic period. Aliasing is also the root of Kepler-82's $3$:$2$ versus $3$:$1$ degeneracy. Hence, even when a fast TTV period is detected, there are often viable solutions that assign it qualitatively different meanings.

KOI-142 is not as vulnerable to this type of ambiguity because the
perturber lies near the $2$:$1$ MMR. At exact $2$:$1$ commensurability, there is one conjunction per orbit of the perturber (i.e., the synodic period is equal to the perturber's period). In such cases, the dominant short-timescale signal has a unique physical interpretation. Although most observed systems are near-resonant, not exactly resonant, this near-equality reduces the susceptibility of $2$:$1$ configurations to alias-based degeneracies.

\subsection{Guidance for TTV solution searches}
\label{sec:TTV_search_advice}

Some practical lessons emerged from our analysis of the 12 TTV planets with reportedly unique solutions:
\begin{itemize}

\item Detection of conjunction-timescale variations appears to be necessary for a unique TTV solution. Without such short-timescale information, long-period TTV signals can be reproduced by perturbers near one of many different MMRs, each corresponding to different planet masses, eccentricities, and mutual inclinations.

\item Detection of conjunction-timescale structure does not guarantee a unique solution. The ability to assign the observed timescale unambiguously to a dynamical timescale is equally important. Because TTVs are measured only once per transit of the perturbed planet, aliasing of the synodic frequency can occur. When a fast TTV signal with period $P_\mathrm{fast}$ is detected in a system with transiting-planet period $P_\mathrm{tra}$, the following possibilities should be examined
carefully:
\begin{itemize}

\item $P_\mathrm{fast}\,{=}\,P_\mathrm{pert}$ (the perturber's orbital period)

\item $P_\mathrm{fast}\,{=}\,P_\mathrm{syn}$, where 
$\frac{1}{P_\mathrm{syn}}\,{=}\,|\frac{1}{P_\mathrm{tra}}\,{-}\,\frac{1}{P_\mathrm{pert}}|$ (the synodic or conjunction period)

\item $P_\mathrm{fast}\,{=}\,P_\mathrm{syn}^{(n)}$,
where $\frac{1}{P^{(n)}_\mathrm{syn}}\,{=}\,
|\frac{1}{P_\mathrm{syn}}\,{-}\,\frac{n}{P_\mathrm{tra}}|$ (alias of
the synodic period)

\end{itemize}

\item Searches employing a fixed period grid, in which the perturber's period is held fixed at one of many different values and the other parameters are optimized, are unlikely to be effective. Because TTV solutions occupy extremely narrow regions of a high-dimensional parameter space, pinpointing a minimum-$\chi^2$ solution can require specification of the perturber's period and eccentricity vector components to five or more decimal places. A brute-force grid spanning plausible ranges of these parameters would entail ${\sim}\,10^{16}$ likelihood evaluations, even before accounting for additional free parameters. Instead, the perturber's period should be allowed to vary from a wide variety of starting conditions.

\item Conventional MCMC methods are likely to struggle with the search, due to the highly multi-modal and sharply defined likelihood surfaces. Nested sampling codes, such as \texttt{MultiNest}, are powerful but not infallible. Our search uncovered viable solutions that were not reported in earlier \texttt{MultiNest}-based searches. It remains possible that our own procedure has missed additional solution families. We are not aware of any general mathematical method that can certify the uniqueness of a TTV solution. Placing an emphasis on perturber periods that seem physically capable of producing the observed TTV timescales and amplitudes can improve search efficiency.

\item When one TTV solution is only marginally preferred over another TTV solution ($\Delta \chi^2\,{\lesssim}\,20$), the accuracy of transit times and uncertainties becomes critical. A small number of timing outliers or modestly underestimated uncertainties can artificially elevate one solution over another and create the false impression of uniqueness.

\end{itemize}

Two future missions, ESA's PLATO mission \citep{Rauer2025} and the China National Space Agency's Earth 2.0 mission \citep{Ge2022}, promise to provide light curves for hundreds of thousands of stars over longer and more continuous time spans than TESS. These missions seem likely to revitalize the transit-timing technique for discovering and characterizing planets. Our study is a reminder that increasing the number of detected TTV systems will not, by itself, provide unambiguous information about planetary perturbers. Accurate timing measurements and uncertainties, thorough parameter searches, and careful dynamical modeling are all essential for reliable inferences about unseen planets.

\section{Acknowledgments} 
\label{sec:acknowledgments}

We are grateful to the authors who have made their transit time and duration measurements publicly available, which helped with our reanalysis. We thank the referee for a thorough report, along with Eric Agol, Sam Hadden, David Nesvorný, Erik Petigura, Shreyas Vissapragada, and Daniel Yahalomi for useful comments and discussions. C.L. acknowledges support from a Natural Sciences and Engineering Research Council of Canada (NSERC) Postgraduate Scholarship.

The data presented in this article were obtained from the Mikulski Archive for Space Telescopes (MAST) at the Space Telescope Science Institute. The observations can be accessed at \dataset[10.17909/95yv-qy54]{https://doi.org/10.17909/95yv-qy54}.

We are pleased to acknowledge that the work reported in this paper was substantially performed using the Princeton Research Computing resources at Princeton University, which is a consortium of groups led by the Princeton Institute for Computational Science and Engineering (PICSciE) and the Office of Information Technology’s Research Computing.

\bibliography{refs}{}
\bibliographystyle{aasjournal}

\end{document}